\documentclass[showpacs,superscriptaddress,amsmath,amssymb,pra]{revtex4}
\usepackage{amssymb}
\usepackage{color}
\usepackage{graphicx}% Include figure files
\usepackage{epsfig}
\usepackage[caption=false]{subfig}
\usepackage{dcolumn}% Align table columns on decimal point
\usepackage{bm}% bold math
\usepackage{CJKutf8}

\newcommand{\beq}{\begin{eqnarray}}
\newcommand{\eeq}{\end{eqnarray}}

\providecommand{\ignore}[1]{}

\begin{document}
\begin{CJK*}{UTF8}{bsmi}

\title{Mechanically Generating Entangled Photons from the Vacuum: A Microwave Circuit-Acoustic Resonator Analogue of the Unruh Effect}

\author{Hui Wang (王惠)}\affiliation{Department of Physics and Astronomy, Dartmouth College, Hanover, New Hampshire
03755, USA }

\author{M. P. Blencowe}\affiliation{Department of Physics and Astronomy, Dartmouth College, Hanover, New Hampshire
03755, USA }

\author{C. M. Wilson}\affiliation{Institute for Quantum Computing and ECE Department, University of Waterloo, Waterloo, Canada}

\author{A. J. Rimberg}\affiliation{Department of Physics and Astronomy, Dartmouth College, Hanover, New Hampshire
03755, USA }

\date{\today}

\begin{abstract}
We consider a model for an oscillatory, relativistic accelerating photodetector inside a cavity and show that the entangled photon pair production from the vacuum (Unruh effect) can be accurately described in the steady state by a non-degenerate parametric amplifier (NDPA), with the detector's accelerating center of mass serving as the parametric drive (pump). We propose an Unruh effect analogue NDPA microwave superconducting circuit scheme, where the breathing mode of the coupling capacitance between the cavity and detector provides the mechanical pump. For realizable circuit parameters, the resulting photon production from the vacuum should be detectable.   
\end{abstract}

\maketitle
\end{CJK*}
\section{\label{sec:intro}Introduction}
In classical physics, the electromagnetic vacuum is devoid of any dynamical activity; accelerating an electrically neutral mirror boundary or accelerating an electromagnetic radiation detector does not result in the generation of energy from the electromagnetic vacuum. In stark contrast, quantum field theory predicts photon pair production from the electromagnetic vacuum for an accelerating mirror boundary, and the detection of photons for an accelerating photodetector. The former process is known as the Dynamical Casimir Effect (DCE), and the latter in the case of a uniformly accelerating detector where the predicted photon spectrum is thermal, the Fulling-Davies-Unruh Effect (or Unruh Effect in short--UE) \cite{Unruh}. 

A longstanding challenge is to demonstrate the DCE and UE in tabletop setups \cite{schutzhold08,schutzhold09}, both to develop a better understanding of photon production from vacuum processes under nonideal experimental situations involving real material systems, and as a possible means for generating entangled photon states directly from the vacuum, providing a resource for quantum information processing applications.

However, a seemingly insurmountable difficulty to demonstrating these effects in the lab is the apparent need to accelerate mechanical mirrors and detector systems up to relativistic speeds in order to get a measurable photon detection signal; in the UE the predicted thermal photon temperature registered by the uniformly accelerating photon detector is $T=\hbar a /2\pi c k_B$, so that detecting $1~{\mathrm{K}}$ thermal photons for example requires a detector proper acceleration $a=2.47\times10^{20}~ {\mathrm{m}}/{\mathrm{s}}^{2}$, which seems impossibly high for any current or planned tabletop experiment.

One approach to circumventing the extreme relativistic speed requirement is to realize analogue systems \cite{nation2012}, which are described by quantum dynamical equations that closely match those for the actual DCE and UE. For example, the accelerating mechanical mirror or photodetector may be replaced by an electromagnetically induced, time changing boundary condition \cite{johansson2009,johansson2010,wilson2011,johansson2013} or detector coupling \cite{delrey12,garcia2017}, such that the quantum electromagnetic field vacuum responds effectively in the same way as in the original DCE and UE. 

In order to see more clearly how such analogues work, consider the situation of an electromagnetic cavity in the single mode approximation for simplicity (we leave until later below a discussion of the validity or otherwise of this approximation). For the DCE, we can suppose that one of the mirrors is oscillating in the longitudinal direction, changing the cavity length. On the other hand, for the UE, we assume that the cavity mirrors are fixed, while we suppose that there is a photon detector with internal degrees of freedom modeled as a harmonic oscillator \cite{lin06,Lin2} and with center of mass oscillating sinusoidally \cite{Doukas} in the longitudinal direction of the cavity; in particular, we relax the usual restricted definition of the UE to allow for non-uniformly accelerating detectors. Under certain approximations, it can be shown that the DCE cavity mode Hamiltonian can then be reduced to that of the degenerate parametric amplifier (DPA):
\begin{equation}
H_{\mathrm{DPA}}=\hbar\omega_c a^{\dag}a +\hbar\lambda \left(e^{-i\Omega_{m}t}a^{\dag\, 2}+e^{i\Omega_{m}t}a^2\right),
\label{DPAeq}
\end{equation}
where $\omega_c$ is the cavity mode frequency, $\Omega_{m}$ is the mirror boundary mechanical center of mass frequency, and the coupling strength parameter $\lambda$ depends on the oscillating mirror amplitude and frequency. On the other hand, the UE effect Hamiltonian can be reduced by approximation to coincide with that of a non degenerate parametric amplifier (NDPA):
\begin{equation}
H_{\mathrm{NDPA}}=\hbar\omega_c a^{\dag}a+\hbar\omega_{d0} b^{\dag}b +\hbar\lambda \left(e^{-i\Omega_{m}t}a^{\dag}b^{\dag}+e^{i\Omega_{m}t}a b\right),
\label{NDPAeq}
\end{equation}
where again $\omega_c$ is the cavity mode frequency, $\omega_{d0}$ is the internal oscillation frequency of the photodetector,  $\Omega_{m}$ is the mechanical center of mass frequency of the photodetector, and $\lambda$ is the coupling strength parameter between the cavity mode and photodetector, which depends on its oscillating center of mass amplitude, frequency, and other electromagnetic factors. Under the resonant drive conditions $\Omega_{m}=2\omega_c$ for the DCE and $\Omega_{m}=\omega_c+\omega_{d0}$ for the UE, Eqs. (\ref{DPAeq}) and (\ref{NDPAeq}) simplify in the interaction picture to
\begin{equation}
H_{\mathrm{DPA}}=\hbar\lambda \left(a^{\dag\, 2}+a^2\right),
\label{DPA2eq}
\end{equation}
\begin{equation}
H_{\mathrm{NDPA}}=\hbar\lambda \left(a^{\dag}b^{\dag}+a b\right),
\label{NDPA2eq}
\end{equation}
and we clearly see that photon pair production in the cavity ($+$ detector) mode(s) occurs, even starting from an initial vacuum state. The analogue then effectively replaces the time dependent mechanical oscillating center of mass frequency terms $e^{\pm i\Omega_{m}t}$ in Eqs. (\ref{DPAeq}) and (\ref{NDPAeq}) with easier to implement oscillating, time dependent electromagnetic terms of identical form \cite{wilson2011,garcia2017}. It is in this way that DCE analogues were realized using superconducting circuit microwave cavity resonators~\cite{wilson2011,lahteen13}, with effective cavity length modulated via oscillating flux tunable Josephson junction inductances.

However, while such  superconducting circuit analogues have experimentally demonstrated photon pair production from vacuum, they are perhaps a little unsatisfying given that non-mechanical parametric amplifiers of both the degenerate and nondegenerate kind are rather ubiquitous in optical and microwave related fields. A more faithful and hence compelling analogue would involve the resonant parametric drive terms $e^{\pm i\Omega_{m}t}$ arising from a genuine {\it mechanically} oscillating system with actual acceleration. For microwave cavity modes in the several GHz regime, where it is straightforward experimentally to cool the modes to close to their quantum vacuum states (i.e., having negligible average thermal photon occupancy number) at mK temperatures, this entails requiring mechanical oscillators with frequencies $\Omega_{m}/(2\pi)\sim 10~{\mathrm{GHz}}$. Such mechanical systems do in fact exist in the form of dilatational or ``breathing" vibrational modes of crystal (or crystalline) solid membrane structures with thicknesses of a few hundred nanometers, and are termed ``film bulk acoustic resonators" (FBAR) when fashioned out of a piezoelectric material that facilitates actuation of the mechanical motion \cite{oconnell2010,cuffe2013}. 

In Ref. \onlinecite{sanz2017}, a superconducting microwave circuit resonator with mechanically oscillating FBAR ``mirror" boundary was proposed as a DCE analogue and it was shown that for realistic system parameters, photon pair production rates from the microwave resonator vacuum should in principle be detectable. Motivated by this proposal, in the present work we analyze a related superconducting circuit scheme incorporating a mechanical FBAR that furnishes a practicable $1+1~{\mathrm{D}}$ Unruh effect analogue. In particular, we show how to realize the NDPA Hamiltonian (\ref{NDPAeq}) with resonant mechanical drive terms.

The in principle detectability can be traced to a number of advantageous features of our scheme. First, by coupling the oscillating photodetector to confined microwave cavity modes as opposed to the unconfined, open space electromagnetic vacuum, the photon pair production rate can be resonantly enhanced \cite{Scully}. This is especially the case for realizable superconducting circuit microwave cavities with confined mode quality factors $Q$ in the tens of thousands and above, and also provided the mechanical FBAR can be actuated in the steady state for multiples of the resonant mode relaxation time $Q/\omega_c$. Coupling a photodetector to a microwave cavity mode necessitates oscillatory center of mass acceleration \cite{Doukas}, rather than the idealized contant proper acceleration considered in the original UE, so that the detector maintains its interaction with the cavity mode. Despite the non-uniform nature of the acceleration, the photodetector nevertheless still ``sees" an effective thermal photon distribution under certain conditions to be established below. This thermal nature follows from the fact that the NDPA Hamiltonian (\ref{NDPA2eq}) generates a two-mode squeezed state starting from vacuum, which appears as a thermal state when either the cavity or photodetector subsystems are traced over.

\begin{figure*}[thb]
\centering
\subfloat[]{\label{udonfig}
  \includegraphics[width=0.4\textwidth]{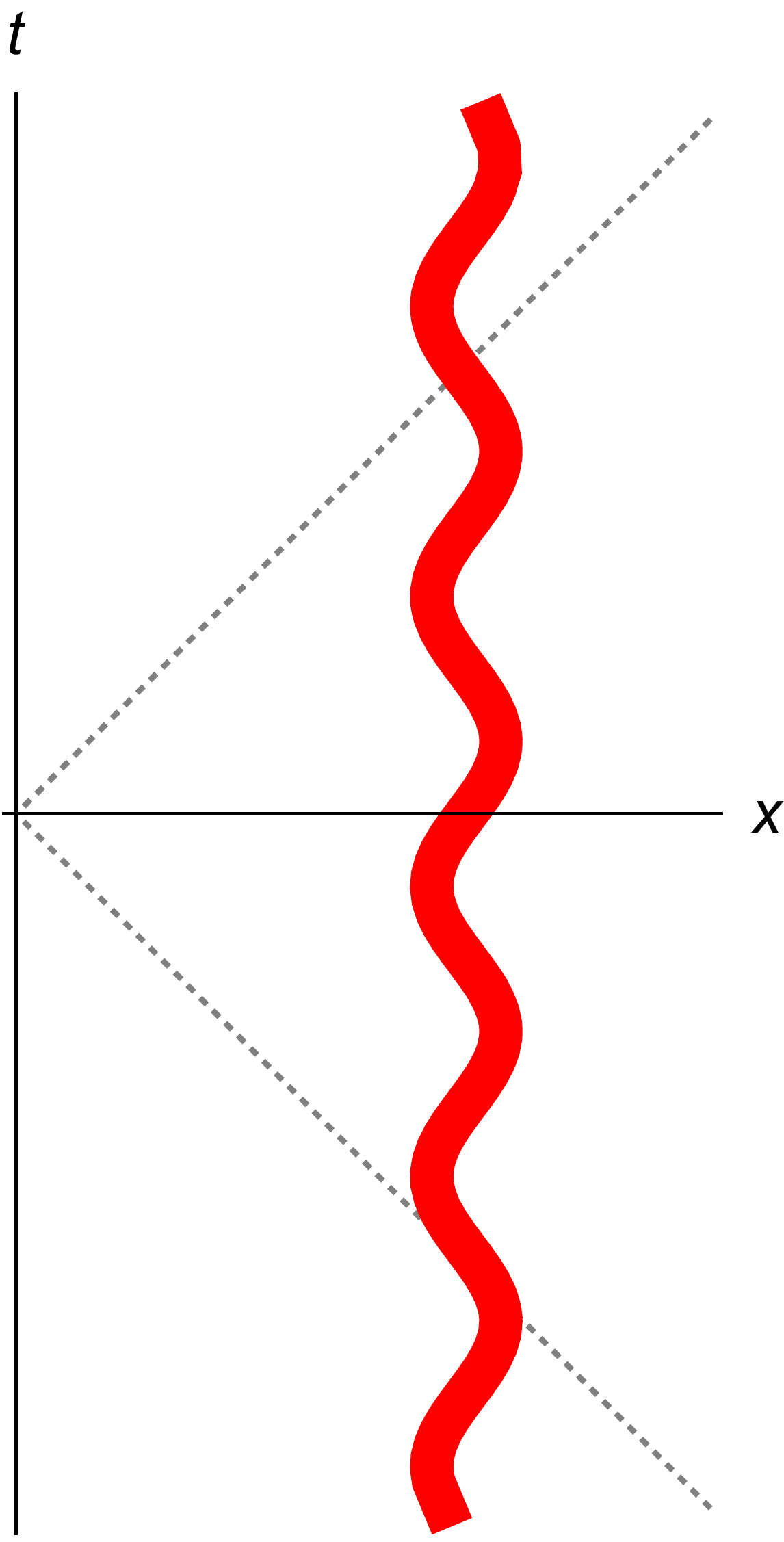}
}
\subfloat[]{\label{sobafig}
  \includegraphics[width=0.4\textwidth]{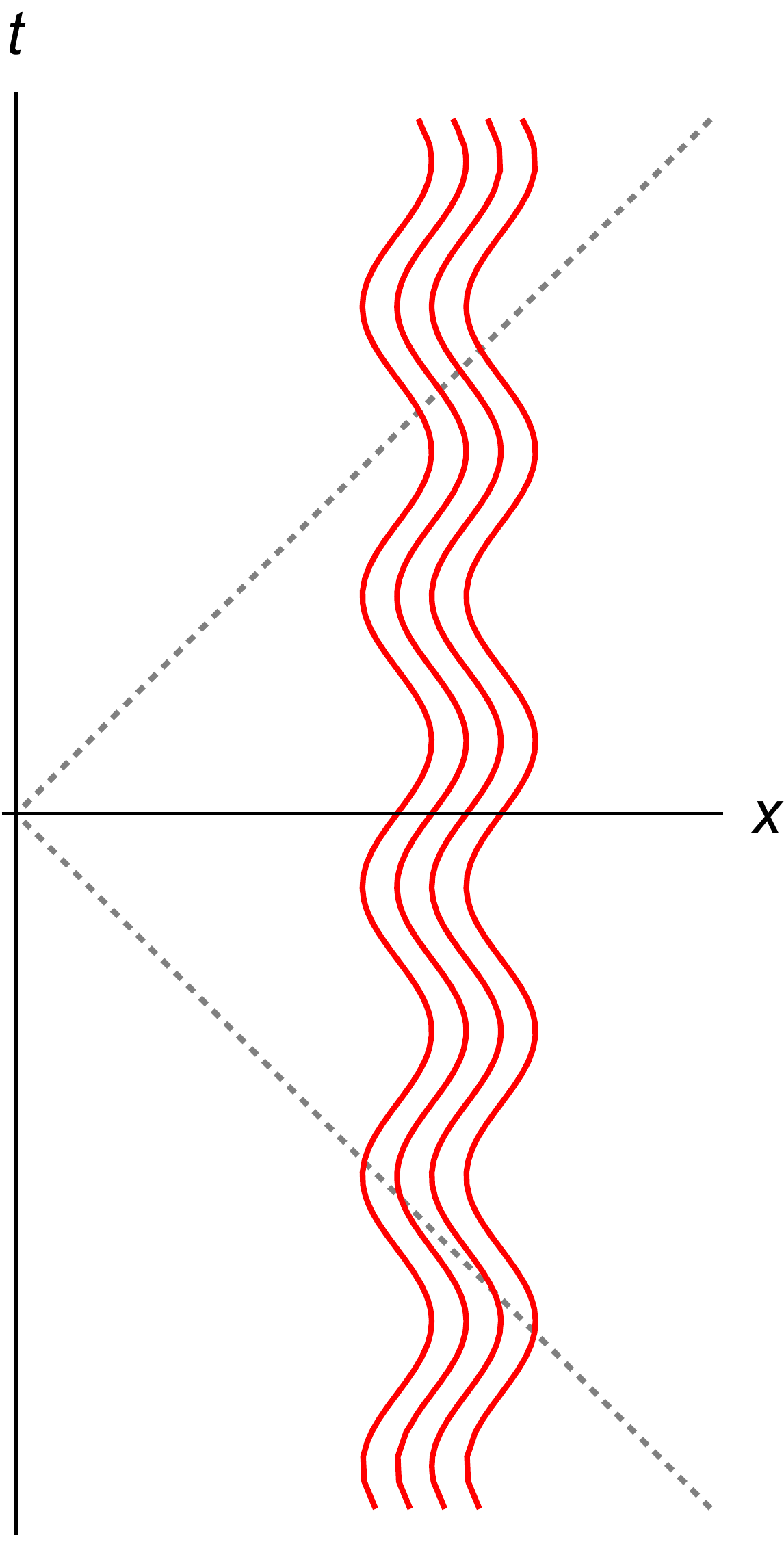}}
\caption{\label{enhancefig} Enhancing the UE with (a) large photodetector crossection, or (b)  many photodetectors.}
\end{figure*} 
Second, the nature of our circuit scheme and that of Ref. \onlinecite{sanz2017} involves utilizing the FBAR also as a capacitor, in our case forming the coupling $\lambda$ between the cavity ($a^{\dag}$, $a$) and internal detector oscillator degrees of freedom ($b^{\dag}$, $b$). By scaling the linear dimension $L_{m}$ of the FBAR capacitance along the cavity resonator length axis, the coupling $\lambda$ correspondingly scales: $\lambda~\sim L_{m}$, and thus the photon pair production rate from vacuum scales as the square of the FBAR linear dimension: $\lambda^2 \sim L_{m}^2$. In this way, the coupling strength can be geometrically enhanced (Fig.~\ref{udonfig}), in contrast to realizations that utilize ``pointlike" photodetectors with dimensions constrained by the microscopic nature of their internal electronic degrees of freedom. In the latter realizations where the photodetector might for example be an atomic scale defect on the surface of the oscillating FBAR, a similar enhancement could instead be achieved through utilizing a large number of defects (Fig.~\ref{sobafig}), all coupled to the same cavity mode. However, as we show later below, increasing the number $N$ of detectors leads to a $\sqrt{N}$ enhancement over the coupling strength for a single defect and hence a photon pair production rate from vacuum that is only linear in $N$. Nevertheless, there is the intriguing possibility of a non-equilibrium, superradiant phase transition \cite{Bastidas},  where the photon pair production rate scales as $N^2$ beyond a critical value of the detector number and coupling, similar to the $L_{m}^2$ scaling of the photon production rate of our present FBAR scheme.

In Sec. \ref{sec:system}, we show that the relativistic, oscillating pointlike photodetector-cavity system action can be reduced by approximation to that of a non degenerate parametric amplifier, even under extreme relativistic, detector center of mass oscillatory acceleration. In Sec. \ref{sec:solving}, we solve for the photodetector-cavity quantum dynamics within the single cavity mode approximation using the Heisenberg-Langevin equation formalism, which takes into account cavity photon loss and accompanying noise. Section \ref{sec:entanglement} examines the effective temperature and entanglement of the photodetector-cavity system in the steady state, and in Sec. \ref{sec:many}, we generalize to many pointlike photodetectors. In Sec. \ref{analogue}, we analyze our superconducting circuit-FBAR analogue of the oscillatory accelerating photodetector-cavity system. Following the conclusion, several appendices provide further details of the analysis. In Appendix \ref{appendixa} we give the derivation of the approximate NDPA Hamiltonian. Appendix \ref{appendixb} derives the second order moments of the cavity-detector quantum dynamics, and Appendix \ref{modemodelsec} gives the derivation of the superconducting circuit-FBAR analogue Hamiltonian and its NDPA approximate form.

\section{\label{sec:system}The Detector-Cavity System}
Our starting UE system comprises  a pointlike detector that is linearly coupled to a $1+1~{\mathrm{D}}$ scalar field denoted as $\Phi(t,x)$ that models the electromagnetic field within the cavity. The detector's center of mass follows the worldline $z^{\mu}(t)=(t,l+A\cos(\Omega_{m} t+\phi))$, where $\Omega_{m}$ is the detector's center of mass oscillation frequency, $\phi$ is a phase constant, $l$ is the average position of the detector center of mass within the $1~{\mathrm{D}}$ cavity, and $A$ is the detector center of mass oscillation amplitude. The detector's internal degrees of freedom are modeled as a quantum harmonic oscillator with rest mass $m_0$, displacement coordinate $Q(\tau)$ and bare natural frequency $\omega_{d0}$. Such a detector model has the advantage that the resulting Hamiltonian is quadratic in the detector-cavity phase space coordinates, so that in the quantum dynamics initial Gaussian states (such as the vacuum state) evolve into Gaussian states that are completely characterized by moments of the coordinate observables only up to second order \cite{lin06,Lin2,Brown}. The action of the combined system is given by \cite{Doukas}:
\begin{equation}
S=-\int d^{1+1}x\frac{1}{2}\partial_{\mu}\Phi\partial^{\mu}\Phi+\int d\tau\left\{\frac{m_0}{2}\left[(\partial_{\tau}Q)^2-\omega_{d0}^2Q^2\right]+\lambda_0\int d^{1+1}xQ(\tau)\Phi(t,x)\delta^{1+1}(x^{\mu}-z^{\mu}(\tau))\right\},
\end{equation}
where $\lambda_0$ is the coupling strength between the detector and cavity field and we adopt the Minkowski metric sign convention $\eta_{\mu\nu}=\mathrm{diag}(-1,1)$. We use $\tau$ to denote the detector's proper time (i.e., time in detector's center of mass rest frame), and $t$
to denote the laboratory (i.e., cavity rest frame) time. With the relation $S=\int dt L$, the system Lagrangian in the laboratory frame is written as
\begin{equation}
L=-\int dx\frac{1}{2}\partial_{\mu}\Phi\partial^{\mu}\Phi+\frac{m_0}{2}\left[\frac{dt}{d\tau}(\partial_{t}Q)^2-\frac{d\tau}{dt}\omega_{d0}^2Q^2\right]+\frac{\lambda_0}{c}\frac{d\tau}{dt}\int dx Q(t)\Phi(t,x)\delta(x-z(t)).
\label{lagrangian}
\end{equation}

We consider a $1~{\mathrm{D}}$ cavity with length $L$ and impose the following Neumann boundary conditions at its $x=0$ and $x=L$ ends: $\Phi^\prime(t,0)=\Phi^\prime(t,L)=0$. Such boundary conditions match the ones that are imposed in the circuit microwave cavity analogue in Sec. \ref{analogue} below, and correspond to the vanishing of the electromagnetic field induced electrical currents at the cavity ends; there would be no essential differences in the following analysis if we were to impose the alternative, Dirichlet boundary conditions $\Phi(t,0)=\Phi(t,L)=0$. The cavity quantum field operator can be decomposed via an expansion in terms of the free field normal mode function solutions:  
\begin{eqnarray}
\Phi(t,x)&=&\sum_n \sqrt{\frac{\hbar c}{n\pi}} \cos\left(k_n x\right) \left(a_n(t)+a_n^{\dag}(t)\right)\cr
&\rightarrow & \sqrt{\frac{\hbar c}{\pi}}\cos\left(k_n x\right)\left(a(t)+a^{\dag}(t)\right),
\label{phi}
\end{eqnarray}
where the free field normal mode wavenumbers are $k_n=\omega_n/c=n\pi/L,\, n=1,2\dots$,  and we introduce photon annihilation/creation operators $a_n$, $a_n^{\dag}$ in a given mode $n$. In the second line of Eq. (\ref{phi}), we truncate the full mode decomposition of the field operator and retain only the lowest, fundamental mode, relabeling the free cavity mode wavenumber/frequency as $k_c=\omega_c/c=\omega_1/c=\pi/L$ and the cavity mode annihilation operator as $a=a_1$; the validity of this single mode approximation will be discussed further below and follows from our restricting to a certain resonance condition between the detector center of mass frequency $\Omega_{m}$, the cavity mode frequency $\omega_{n=1}$, and the center of mass rest frame frequency $\omega_{d0}$ of the detector's internal degree of freedom. The truncated momentum operator $\Pi(t,x)$ that is conjugate to the truncated field operator expression (\ref{phi}) for $\Phi(t,x)$ is given by 
\begin{equation}
\Pi(t,x)=-\frac{i}{L}\sqrt{\frac{\hbar\pi}{c}}\cos(k_c x)\left(a(t)-a^{\dag}(t)\right),
\end{equation}
The internal detector oscillator position and momentum operators in terms of the annihilation/creation operators $b$ and $b^{\dag}$ are respectively
\begin{eqnarray}
Q&=&\sqrt{\frac{\hbar}{2\omega_{d0}m_0}}\left(b+b^{\dag}\right)\cr
P&=&-i\sqrt{\frac{\hbar\omega_{d0}m_0}{2}}\left(b-b^{\dag}\right).
\label{internaleq}
\end{eqnarray}
Performing the Legendre transformation on the Lagrangian  (\ref{lagrangian}) and replacing the position and momentum coordinates with their corresponding operators, we obtain the following cavity-detector system quantum Hamiltonian expressed in dimensionless units:
\begin{equation}
\tilde{H}(\tilde{t})=H(t)/(\hbar\omega_c)=a^{\dag}a+\tilde{\omega}_{d0}\frac{d\tau}{dt}b^{\dag}b - \tilde{\lambda}_0\frac{d\tau}{dt} \cos \left[ k_c l+ k_c A\cos\left(\tilde{\Omega}_{m} \tilde{t}+\phi\right)\right]\left(a^{\dag}+a\right)\left(b^{\dag}+b\right),
\label{orihamiltonian}
\end{equation}
where we have introduced the dimensionless time coordinate $\tilde{t}=\omega_c t$, such that $\tilde{\omega}_{d0}=\omega_{d0}/\omega_c$ and $\tilde{\Omega}_{m}=\Omega_{m}/\omega_c$.
The dimensionless cavity mode-detector coupling strength is $\tilde{\lambda}_0=\lambda_0/(\omega_c\sqrt{2\pi m_0\omega_{d 0} c})$. The Lorentz factor $d\tau/{dt}=\sqrt{1-\xi^2\sin^2(\Omega_{m} t+\phi)}$ accounts for the redshift in the detector oscillator frequency $\tilde{\omega}_{d0}$ as measured by an observer in the laboratory frame, with $\xi=\Omega_{m} A/c$ the ratio of the detector center of mass velocity magnitude to the speed of light $c$; note that the factor $k_c A$ appearing in the mode function cosine argument can also be expressed as $\xi/\tilde{\Omega}_{m}$.

We drop the tildes from now on for notational convenience and locate the detector's center of mass equilibrium position at the midpoint $l=L/2$ of the cavity, i.e., at the node of the fundamental normal mode function where its gradient and hence cavity-detector coupling is a maximum. The time-dependent part of the interaction term in Eq.~(\ref{orihamiltonian}) then reduces to
\begin{equation}
\frac{d\tau}{dt}\cos \left[k_cl+\frac{\xi }{\Omega_{m}}\cos(\Omega_{m} t)\right]= -\frac{d\tau}{dt}\sin \left[\frac{\xi}{\Omega_{m}}\cos(\Omega_{m}t) \right],
\end{equation}
where we have set $\phi=0$ since the phase does not affect the dynamical behavior in any essential way.

Given that the time dependent Hamiltonian (\ref{orihamiltonian}) is periodic with period $2\pi/\Omega_{m}$, we can approximate the Hamiltonian as a series in harmonics of $\Omega_{m}$ via a Fourier expansion of the Lorentz factor $d\tau/dt$, and a Jacobi-Anger expansion of the sinusoidal term $\sin[\xi\cos(\Omega_{m}t)/\Omega_{m}]$. The Hamiltonian then becomes up to second harmonic, time-dependent terms  $e^{\pm 2i\Omega_mt}$ (see Appendix ~\ref{appendixa}):
\begin{equation}
H=a^{\dag}a+\omega_{d0}\left[D_0+D_2\cos(2\Omega_mt)\right]b^{\dag}b+\lambda_0 C_1\cos(\Omega_mt)(a^{\dag}+a)(b^{\dag}+b),
\label{reduced}
\end{equation}
where $D_0$ and $D_2$ are coefficients that depend on $\xi$, while the coefficient $C_1$ depends on $\xi$ and $\Omega_m$. 
We now impose the resonance condition $\Omega_m=1+\omega_{d}$, i.e., the oscillating detector center of mass frequency matches the sum of the optical cavity frequency and renormalized detector frequency $\omega_d=\omega_{d0} D_0$.
Transforming to the rotating frame via the unitary operator $U_{\mathrm{RF}}(t)=\exp\left({ia^{\dag}at}+{ib^{\dag}b\left[\omega_dt+\frac{\omega_{d0}D_2}{2\Omega_m}\sin(2\Omega_mt)\right]}\right)$ and performing the rotating wave approximation (RWA) by dropping rapidly oscillating terms, we obtain the following time-independent Hamiltonian in the interaction picture:
\begin{equation}
H_I=\lambda(a^{\dag}b^{\dag}+ab),
\label{rwa}
\end{equation}
where the renormalized coupling is $\lambda=\frac{1}{2}{\lambda_0 C_1}\left[J_0\left(\frac{\omega_{d0}D_2}{2\Omega_m}\right)-J_1\left(\frac{\omega_{d0}D_2}{2\Omega_m}\right)\right]$, with $J_0(z)$ and $J_1(z)$ Bessel functions of the first kind. We recognize Hamiltonian (\ref{rwa}) to be that of a non-degenerate parametric amplifier (NDPA), where the detector's center of mass mechanical motion plays the role of the pump, while the detector and cavity modes play the role of the so-called signal and idler.  Assuming that the detector and cavity modes are initially in their vacuum state, Eq.~(\ref{rwa}) describes the parametric amplification of the vacuum fluctuations (i.e., photon pair production), resulting in a two mode squeezed state.

\section{\label{sec:solving}Solving for the Quantum Dynamical Evolution}
In actual realizations, the cavity mode and detector oscillator subsystems will be open, interacting with external environmental degrees of freedom. This results in damping and associated noise forces acting on these systems. The input-output formulation \cite{gardiner1985,gardiner2000} provides a convenient approach to incorporating damping and noise, and together with the closed system dynamics following from the approximated Hamiltonian (\ref{rwa}), results in the following quantum Langevin equations for the cavity mode and detector lowering and raising operators $a$, $a^{\dag}$, $b$ and $b^{\dag}$:
\begin{equation}
\frac{d}{dt}
\begin{pmatrix}
a \\ a^{\dagger} \\ b \\ b^{\dagger}\\
\end{pmatrix}
=
\begin{pmatrix}
-\frac{\gamma}{2} &0 &0 &i\lambda\\
0 &-\frac{\gamma}{2} &-i\lambda &0\\
0 &i\lambda &-\frac{\gamma}{2} &0\\
-i\lambda &0 &0 &-\frac{\gamma}{2} \\
\end{pmatrix}
\begin{pmatrix}
a \\ a^{\dagger} \\ b \\ b^{\dagger}\\
\end{pmatrix}
+\sqrt{\gamma}
\begin{pmatrix}
a_{\mathrm{in}} \\ a_{\mathrm{in}}^{\dagger} \\ b_{\mathrm{in}} \\ b_{\mathrm{in}}^{\dagger}\\
\end{pmatrix},
\label{ode}
\end{equation}
where the cavity mode and detector oscillator are assumed here for simplicity to have the same energy damping rate $\gamma$. We shall consider the weakly damped regime for the cavity and detector oscillator, i.e., $\gamma\ll\omega_c,\,\omega_d$, since this favors enhanced photon production rates from the vacuum. The input noise operators $a_\mathrm{in}(t)$, $b_\mathrm{in}(t)$ satisfy the expectation value and correlation relations $\langle a_{\mathrm{in}}(t)\rangle=0$, $\langle b_{\mathrm{in}}(t)\rangle=0$ $\langle a_{\mathrm{in}}(t) a_{\mathrm{in}}(t')\rangle=0$, $\langle b_{\mathrm{in}}(t) b_{\mathrm{in}}(t')\rangle=0$, $\langle a^{\dag}_{\mathrm{in}}(t) a_{\mathrm{in}}(t')\rangle=0$, $\langle b^{\dag}_{\mathrm{in}}(t) b_{\mathrm{in}}(t')\rangle=0$  and  $\langle a_{\mathrm{in}}(t)a_{\mathrm{in}}^{\dag}(t^{\prime})\rangle=\delta(t-t^{\prime})$, $\langle b_{\mathrm{in}}(t)b_{\mathrm{in}}^{\dag}(t^{\prime})\rangle=\delta(t-t^{\prime})$, where  we assume the environment temperature to be negligible compared to the frequencies of the cavity mode and detector oscillator (i.e., $k_B T\ll \hbar\omega_c,\, \hbar\omega_d$). We furthermore assume that the cavity mode and detector oscillator noise operators are uncorrelated, i.e., $\langle b_{\mathrm{in}}(t)a_{\mathrm{in}}^{\dag}(t^{\prime})\rangle=0$.   

Suppose that the detector and cavity mode oscillators are initially ($t=0$) in their ground states. Since both the full and approximated  Hamiltonians, Eqs.  (\ref{orihamiltonian}) and (\ref{rwa}) respectively, are quadratic in the bosonic operator terms, such an initial Gaussian state remains Gaussian throughout its evolution (also with damping included); the first order moments vanish, so that the second order moments in $a$ and $b$ and their conjugates therefore completely determine the cavity mode-detector state. With both cavity mode and detector oscillator subject to damping and noise, in the long time limit they will reach a steady state that is independent of their initial state. The analytical solutions for the nonzero steady state second moments that follow from Eqs. (\ref{ode}) are 
\begin{equation}
\left.\langle a^{\dag}(t)a(t)\rangle\right|_{t\rightarrow\infty} =\left.\langle b^{\dag}(t)b(t)\rangle\right|_{t\rightarrow\infty}=\frac{2\eta^2}{1-4\eta^2},
\label{secondmoments1}
\end{equation}
\begin{equation}
\left.\langle a^{\dag}(t)b^{\dag}(t)\rangle\right|_{t\rightarrow\infty} =-\left.\langle a(t)b(t)\rangle\right|_{t\rightarrow\infty}=\frac{i\eta}{1-4\eta^2}.
\label{secondmoments2}
\end{equation}
The moments in Eq. (\ref{secondmoments1}) are the steady state average photon numbers of the cavity mode and detector oscillator.  
Note that, by virtue of Eqs.~(\ref{secondmoments1}) and ~(\ref{secondmoments2}), the state of the system is
entirely governed by the dimensionless parameter $\eta=\lambda/\gamma$ which expresses the ratio of the coupling strength $\lambda$ to the damping rate $\gamma$. For stable steady state solutions to exist,  we require $\eta<\eta_\mathrm{crit}=1/2$; as the critical value $\eta_\mathrm{crit}=1/2$ is approached from below, all four non-zero second moments (and in particular the average photon numbers) approach infinity, with the evolution time required to reach the steady state also approaching infinity. Beyond this critical value, the system exhibits the so-called parametric instability. Such an instability is a consequence of assuming a harmonic oscillator internal detector degree of freedom; for a more realistic model of a detector involving an {\emph{anharmonic}} oscillator (which is commonly approximated by a truncated two level system), such an instability does not arise, although the dynamics is no longer solvable analytically. 
 
\begin{figure*}[thb]
\centering
\subfloat[]{\label{expectationa}
  \includegraphics[width=0.45\textwidth]{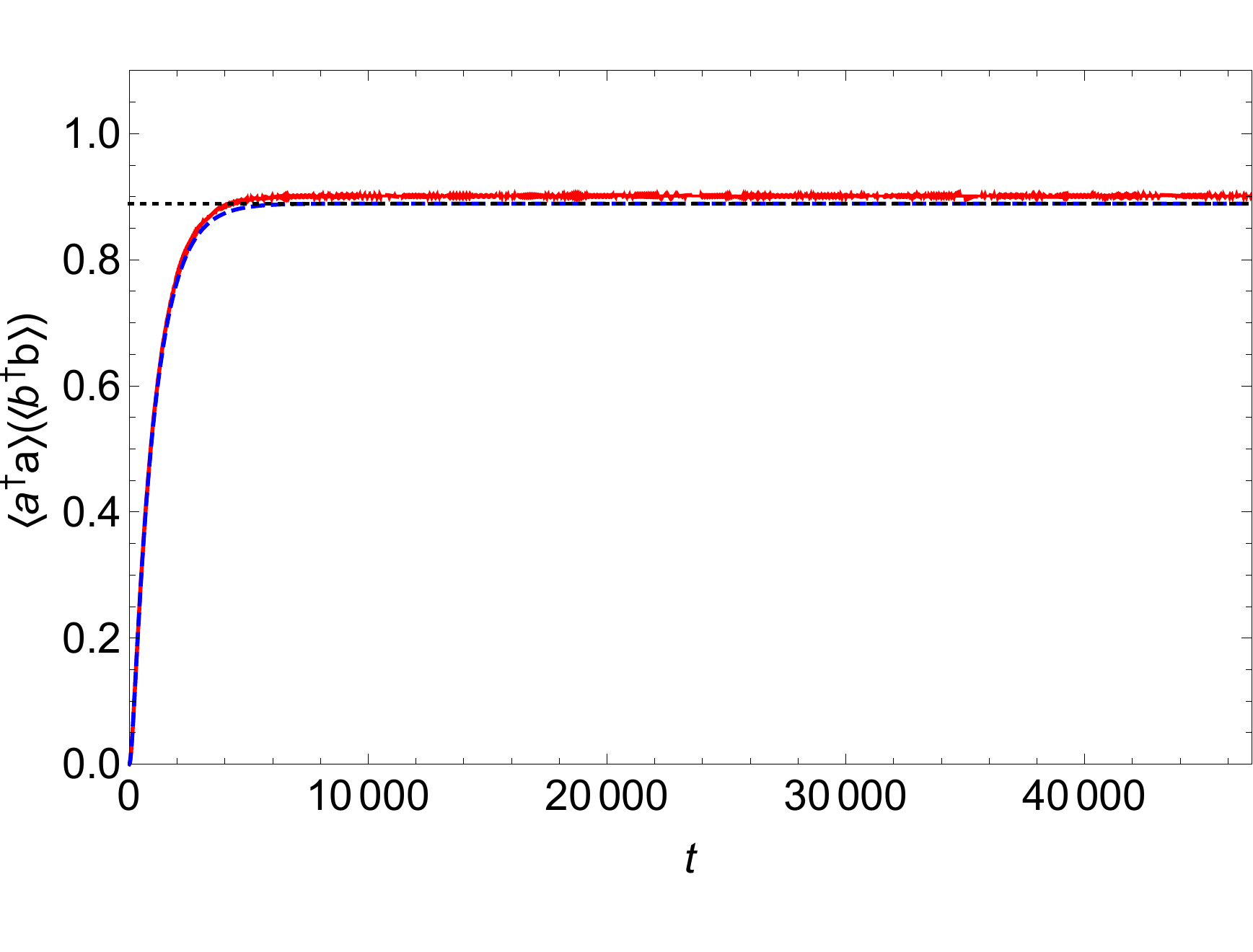}
}
\subfloat[]{\label{expectationb}
  \includegraphics[width=0.45\textwidth]{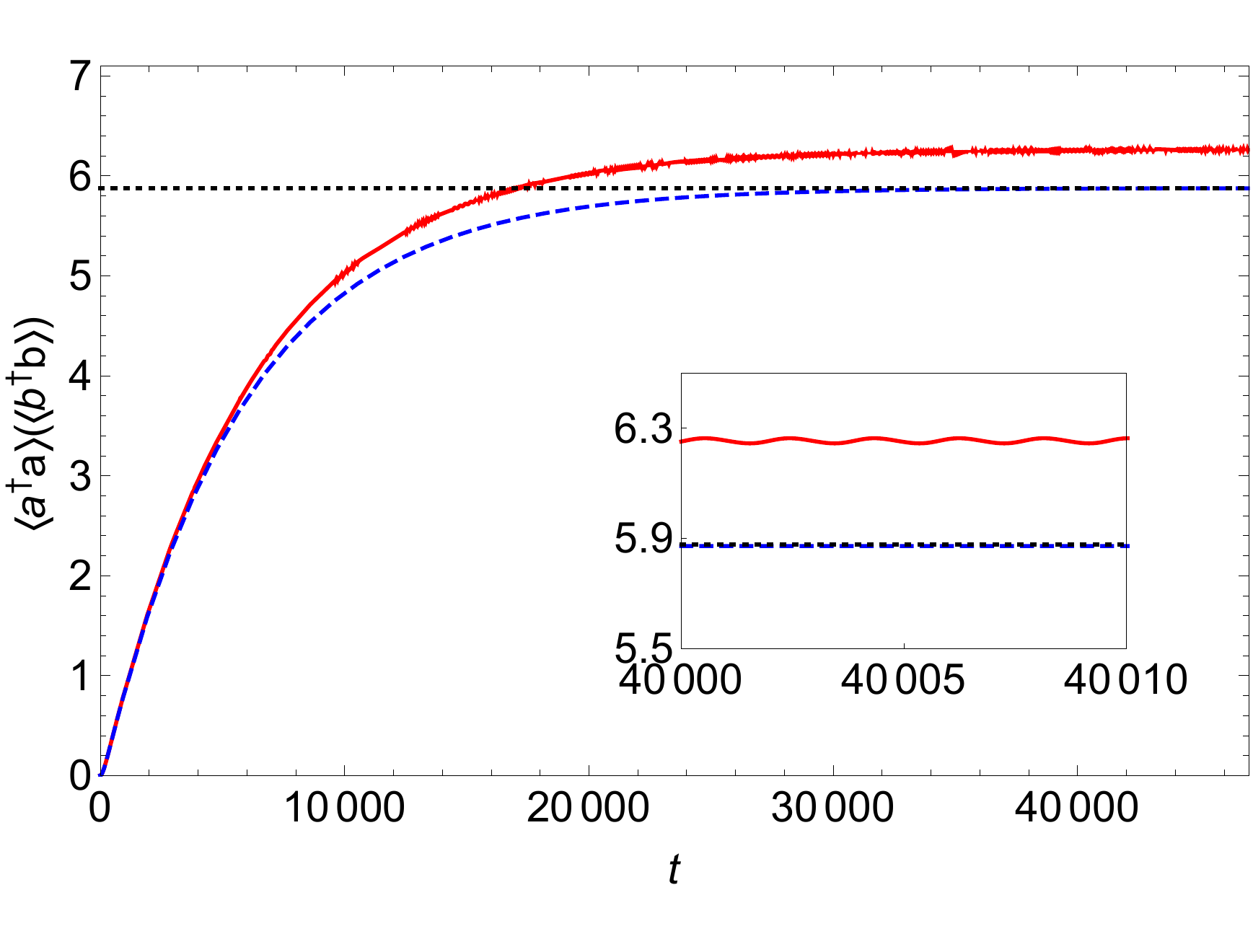}}

\caption{Average photon number of the cavity mode and detector oscillator versus time for the numerical solution to th quantum Langevin equation with  full Hamiltonian (\ref{orihamiltonian})  (solid curve), the analytical solution to the quantum Langevin equation with the RWA Hamiltonian [Eqs. (\ref{ode}) and (\ref{tdepsecondmom1eq})] (Dashed), and the steady state solution Eq.~(\ref{secondmoments1}) (dotted curve). The parameter values are $\xi=0.8$, $\omega_{d 0}=0.8$ with renormalized $\omega_d=0.65$, $\Omega_m=1.65$, and  $\lambda_0=0.01$  with renormalized $\lambda=0.0021$. (a) $\eta=0.40$ corresponding to $\gamma\approx 0.005$; (b) $\eta=0.48$ corresponding to $\gamma\approx 0.004$.}
\label{expectation}
\end{figure*} 

In Fig.~\ref{expectation}, we consider some example parameter values such that the resonance condition holds ($\Omega_m=1+\omega_d$), and compare the photon number expectation values for the cavity mode and detector oscillator obtained by numerically solving the quantum Langevin equation with the full Hamiltonian (\ref{orihamiltonian}), the analytical solution to the quantum Langevin equation with the RWA Hamiltonian [Eqs. (\ref{ode}) and (\ref{tdepsecondmom1eq})], and the analytical steady state solution (\ref{secondmoments1}). Note that the cavity and detector photon number expectation values are identical for the RWA Hamiltonian  description [see Eq. (\ref{secondmoments1})], while they closely coincide for the full Hamiltonian description. This is a consequence of the fact that the photons are produced in pairs (one in the cavity and one in the detector) starting from the system vacuum state, and that we assume the same damping rates for the cavity mode and detector (given by $\gamma$). Note also that in this example, we consider quite an extreme relativistic velocity magnitude: $\xi=\Omega_mA/c=0.8$; nevertheless, the quantum Langevin equation with simple NDPA Hamiltonian (\ref{rwa}) and appropriately renormalized coupling $\lambda$ still accurately describes the average photon numbers. Comparing Figs. \ref{expectationa} and \ref{expectationb} which correspond to different $\eta=\lambda/\gamma$ values, we see that as $\eta$ approaches the parametric instability threshold $\eta_{\mathrm{crit}}=1/2$ the average photon numbers of the cavity mode and detector increase (i.e., diverge), while the evolution time required to reach the steady state also increases. Furthermore, note that the deviation between the results obtained using the full Hamiltonian and the RWA Hamiltonian grows as $\eta$ approaches the instability threshold. The  Fig.~\ref{expectationb} inset provides a zoom-in view of the average photon numbers in the long-time limit; the oscillatory behavior of the full Hamiltonian dynamics arises from time-dependent oscillatory terms which are neglected in the RWA Hamiltonian [see Eq. (\ref{interaction2})]. The deviation in the average photon numbers between the full and RWA descriptions is a consequence of the fact that small errors in determining the renormalized coupling $\lambda$ (due to neglecting higher than 2nd harmonics) get amplified as $\eta$ approaches the instability threshold.

In solving for the quantum dynamics, we have neglected the higher cavity modes $\omega_n=n\pi c/L=n\omega_c =n\geq 2$ (in dimensionless units). This is justified in the steady state provided the higher frequency harmonics $k\Omega_m,\, k\geq 2$, of the center of mass drive are not resonant with the sum of a higher cavity mode $+$ detector frequency: $n+\omega_d$. In particular, we require that  
\begin{equation}
\left|k\Omega_m-n -\omega_d\right|\gg \gamma.
\label{singlemodeeq}
\end{equation}
Substituting in the resonance condition $\Omega_m=1+\omega_d$, we obtain
\begin{equation}
\left|(k-n) +(k-1)\omega_d\right|\gg \gamma,
\label{singlemode2eq}
\end{equation}
which for the above example parameter values becomes
\begin{equation}
\left|(k-n) +0.65 (k-1)\right|\gg 0.005
\label{singlemode3eq}
\end{equation}
when, e.g., $\eta=0.40$. Condition (\ref{singlemode3eq}) is not violated until we go up to $n=34$ and $k=21$. However, the effective coupling at such a high harmonic is much smaller than that for $n=k=1$, so that the photon production in the higher mode can be neglected. In essence, higher cavity modes can be neglected provided the ratio of the detector to fundamental cavity frequency $\omega_d$ is not an irreducible fraction with small numbers in the numerator and denominator.  

On the other hand, during the initial transient, evolving stage after the center of mass oscillator drive is ``switched on", we do expect higher cavity modes to be populated with generated photons, especially for the example extreme relativistic velocity magnitude $\xi=0.8$ considered above \cite{Loop}. Thus, the time evolution given in Fig. \ref{expectation} is likely to be accurate only during the steady state regime. For the analogue circuit realization considered below in Sec. \ref{analogue}, the single cavity mode approximation should also be accurate in the transient regime since in that case we have $\xi\lll 1$.

\section{\label{sec:entanglement} Effective Temperature and Entanglement}
\begin{figure*}[thb]
\centering
\subfloat[]{\label{tempefig}
  \includegraphics[width=0.45\textwidth]{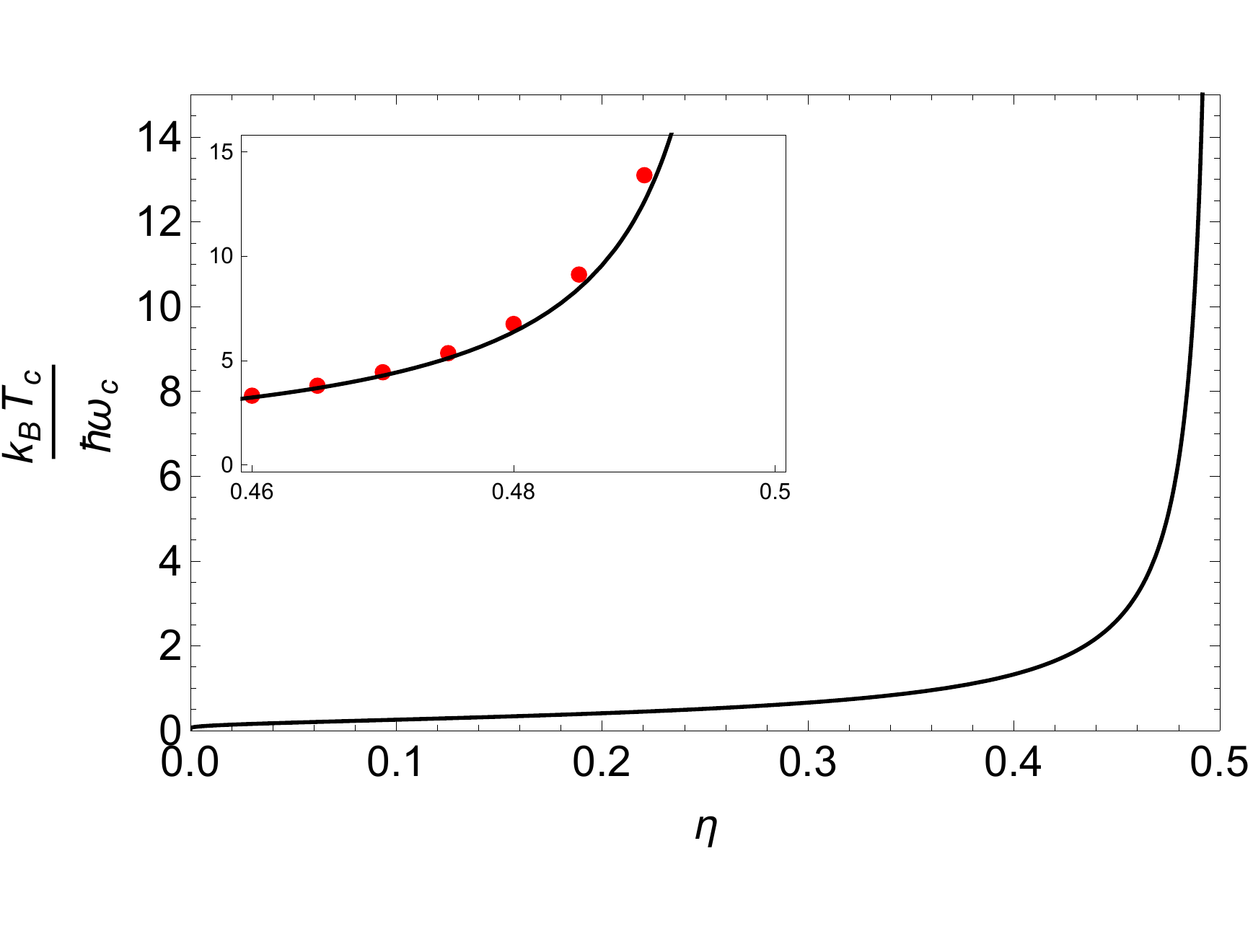}
}
\subfloat[]{\label{entanglfig}
  \includegraphics[width=0.45\textwidth]{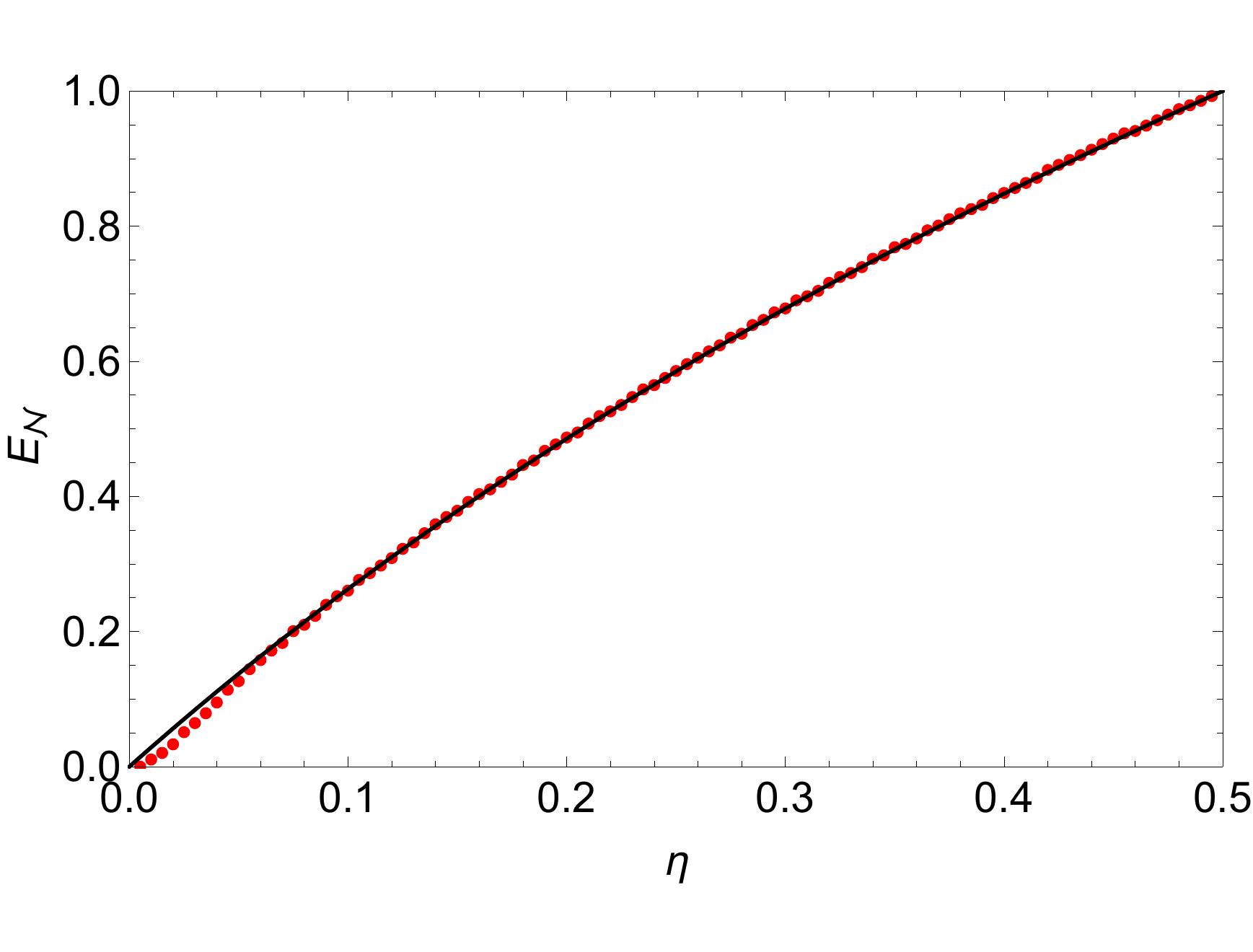}}
\caption{\label{tempentanglefig} Steady state value of (a) cavity mode oscillator effective temperature as a function of $\eta=\lambda/\gamma$ given by the RWA analytical formula (\ref{cavteq}). The inset plot compares the analytical formula (solid line) with the effective temperature obtained using the full Hamiltonian (\ref{orihamiltonian}) (dots).  (b) Entanglement (logarithmic negativity $E_{\mathcal{N}}$) as a function of $\eta$ given by the RWA analytical formula (\ref{entanglementeq}) (solid line) and by the full Hamiltonian (\ref{orihamiltonian}) (dots) ; $E_{\mathcal{N}}$ ranges between $0$ (for $\eta=0$) and $1$ (for $\eta=\eta_{\mathrm{crit}}=1/2$).}
\end{figure*}

Due to the fact that the system state is Gaussian and the first order moments vanish, the state is completely determined by the covariance matrix $\Gamma$ with elements $\Gamma_{\alpha\beta}=\langle R_{\alpha}R_{\beta}+R_{\beta}R_{\alpha}\rangle/2$ where $\vec{R}=(X_a,P_a,X_b,P_b)$ are the quadrature amplitudes of the cavity and detector modes, related to the creation/annihilation operators as follows:
\begin{equation}
\begin{pmatrix}
X_a \\ P_a \\ X_b \\ P_b\\
\end{pmatrix}
=M
\begin{pmatrix}
a \\ a^{\dagger} \\ b \\ b^{\dagger}\\
\end{pmatrix},
M
=\frac{1}{\sqrt{2}}
\begin{pmatrix}
1 &1 &0 &0\\
-i &i &0 &0\\
0 &0 &1 &1\\
0 &0 &-i &i \\
\end{pmatrix}.
\label{pmatrixeq}
\end{equation}
An oscillator thermal state is characterized by a zero-mean, circularly-symmetric Gaussian Wigner function distribution on phase space. For the NDPA Hamiltonian (\ref{rwa}), the fact that $\langle a(t)\rangle$, $\langle a(t)^2\rangle$, $\langle b(t)\rangle$ and $\langle b(t)^2\rangle$ vanish throughout the evolution implies that $\langle R_{\alpha}\rangle$ vanishes, so that we have $\langle X_a^2\rangle=\langle P_a^2\rangle$ and $\langle X_b^2\rangle=\langle P_b^2\rangle$. Thus, the reduced states of both the cavity mode and detector oscillator are exact thermal states within the RWA Hamiltonian approximation. Numerical solutions of the quantum Langevin equation for the full Hamiltonian (\ref{orihamiltonian}) give  $|\langle a(t)^2\rangle|\ll \langle a^{\dag}(t) a(t)\rangle$ and $|\langle b(t)^2\rangle|\ll \langle b^{\dag}(t) b(t)\rangle$ in the steady state for $0<\eta<1/2$, so that the cavity mode and detector states are thermal (i.e., described by a Boltzmann distribution) to a good approximation.  For an arbitrary single bosonic mode with frequency $\omega$ in a thermal state defined by temperature $T$, the expectation value of the mode occupation number is given by the Bose-Einstein distribution $\langle N \rangle=1/(e^{\hbar\omega/k_BT}-1)$. Inverting and substituting in Eq. (\ref{secondmoments1}), we obtain the following expressions for the approximate effective temperatures of the cavity mode and detector:
\begin{eqnarray}
\frac{k_B T_c(t)}{\hbar\omega_c} &=&\frac{1}{\ln\left(\frac{\langle a(t)^{\dag}a(t)\rangle+1}{\langle a(t)^{\dag}a(t)\rangle}\right)}\xrightarrow[]{t\rightarrow\infty} \frac{1}{\ln\left(\frac{1-2\eta^2}{2\eta^2}\right)},\label{cavteq}\\
\frac{k_B T_d(t)}{\hbar\omega_d} &=&\frac{1+D_2/D_0\cos(2\Omega_mt)}{\ln\left(\frac{\langle b(t)^{\dag}b(t)\rangle+1}{\langle b(t)^{\dag}b(t)\rangle}\right)}\xrightarrow[]{t\rightarrow\infty}\frac{1+D_2/D_0\cos(2\Omega_mt)}{\ln\left(\frac{1-2\eta^2}{2\eta^2}\right)}.\label{detteq}
\end{eqnarray}
The time dependence of the detector effective temperature $T_d$ in the steady state  arises from the time-varying Lorentz factor $d\tau/dt\approx D_0+D_2\cos(2\Omega_mt)$, which ``red shifts" the frequency of the detector's internal mode as viewed from the lab frame--a consequence of the detector's center of mass oscillatory motion. (The coefficients $D_0$, $D_2$ are discussed in Appendix ~\ref{appendixa}.) The effective temperature of the cavity mode oscillator is shown in Fig.~\ref{tempefig}; the temperature factor $k_B T_c/\hbar\omega_c$ behaves like $1/[4(1-2\eta)]$, which increases (i.e., diverges) as $\eta$ approaches the instability threshold value $1/2$. The inset of Fig.~\ref{tempefig} compares the analytical approximation (\ref{cavteq}) with the full numerical calculation of the effective temperature; the discrepancy between the two calculations grows as $\eta$ approaches the instability threshold, again a consequence of small errors in determining the renormalized coupling $\lambda$ (due to neglecting higher than 2nd harmonics) getting amplified.     

Starting from their initial ground states, the NDPA Hamiltonian generates a quantum entangled state between the cavity mode and detector oscillator. In the presence of environmental dissipation and noise, the system state is a mixed Gaussian state, so that an appropriate entanglement measure is the so-called logarithmic negativity $E_\mathcal{N}$, which derives from the positive partial transpose criterion for a separable state \cite{Vidal,Wolf}. The partial transpose operation corresponds to switching the sign of the quadrature degree of freedom of one of the two oscillators; switching the sign of $X_a$ by making transformation $R\to\Lambda R$, where $\Lambda=\mathrm{diag}(-1,1,1,1)$, the partially transposed matrix becomes
\begin{equation}
\Gamma^{PT}
=\Lambda M
\begin{pmatrix}
\langle a^2\rangle & \langle a^{\dag}a\rangle+\frac{1}{2} & \langle ab\rangle & \langle ab^{\dag}\rangle\\ 
\langle a^{\dag}a\rangle+\frac{1}{2} & \langle a^{\dag 2}\rangle & \langle a^{\dag}b\rangle & \langle a^{\dag}b^{\dag}\rangle \\
\langle ab\rangle & \langle a^{\dag}b\rangle & \langle b^2\rangle & \langle b^{\dag}b\rangle+\frac{1}{2}\\ 
\langle ab^{\dag}\rangle & \langle a^{\dag}b^{\dag}\rangle & \langle b^{\dag}b\rangle+\frac{1}{2} & \langle b^{\dag2}\rangle\\
\end{pmatrix}
M^T\Lambda.
\label{transposed}
\end{equation}
Substituting Eqs.~(\ref{secondmoments1}) and ~(\ref{secondmoments2}) into  Eq.~(\ref{transposed}) with all other moments set to zero, we find that $\Gamma^{PT}$ has the following two 
eigenvalues: $e_1=1/{[2(1+2\eta)]},e_2=1/{[2(1-2\eta)]}$. In terms of these eigenvalues, the entanglement in the long time limit as defined by the logarithmic negativity is
\begin{equation}
E_{\mathcal{N}}=\sum_{i=1,2}\max[0,-\log_2(2e_i)]=\log_2(1+2\eta).
\label{entanglementeq}
\end{equation}
In Fig.~\ref{entanglfig} we see that the entanglement increases monotonically from $0$ with increasing $\eta$ and reaches its maximum value $E_\mathcal{N}^{\max}=1$ at $\eta_{\mathrm{crit}}=1/2$; the  RWA formula (\ref{entanglementeq}) accurately matches the logarithmic negativity obtained numerically from the quantum Langevin equation for the full Hamiltonian (\ref{orihamiltonian}) thoughout the $\eta$ range, apart from small $\eta\rightarrow 0$ values.     

\section{\label{sec:many}Many detectors}

A single accelerating detector coupled to the cavity mode generates photons by extracting energy from the detector center of mass motion. If more than one accelerating detector is brought into play, coupling to the same cavity mode, we might expect an enhancement of the Unruh effect. Consider $N$ detectors with their center of mass equilibrium points located in the region of the midway point of the cavity with the spatial separation between the detectors much smaller than  the cavity fundamental mode wavelength $2L$; the detectors approximately follow the worldlines $z^{\mu}_n(t)=(t,L/2+A\cos(\Omega_m t+\phi_n))$. Here we allow for the possibility that the detectors have different phases $\phi_n$, but with identical center of mass oscillation amplitudes $A$ and frequencies $\Omega_m$. Assuming for simplicity that the coupling strengths between each detector and the cavity mode are given by the same $\lambda_0$, the single detector Hamiltonian (\ref{orihamiltonian}) is replaced by
\begin{equation}
H(t)=a^{\dag}a+\omega_{d0}\sum_{n=1}^N\frac{d\tau_n}{dt}b_n^{\dag}b_n+ \lambda_0\sum_{n=1}^N\frac{d\tau_n}{dt}\sin\left[k_cA\cos(\Omega_m t+\phi_n)\right](a^{\dag}+a)(b_n^{\dag}+b_n).
\label{manydetectors}
\end{equation}
Following the same approximation procedure (harmonic series expansion and RWA) as for the single detector case (see Appendix ~\ref{appendixa}), Hamiltonian (\ref{manydetectors}) can be approximated by the following time-independent Hamiltonian in the interaction picture:
\begin{eqnarray}
H_I & \approx & \lambda\sum_{n=1}^N(a^{\dag}b_n^{\dag}+ab_n),
\label{approxhamil}
\end{eqnarray}
where the renormalized coupling is given by the same expression as in the single detector case. Consider the collective detector operator $\displaystyle b_{\mathrm{col}}=\frac{1}{\sqrt{N}}\sum_{n=1}^Nb_n$ and $\displaystyle b_{\mathrm{in},\mathrm{col}}=\frac{1}{\sqrt{N}}\sum_{n=1}^Nb_{\mathrm{in},n}$, where the $1/\sqrt{N}$ factors ensure that the usual commutation relations   are obeyed: $\left[b_{\mathrm{col}},b_{\mathrm{col}}^{\dag}\right]=1$ and $\left[b_{\mathrm{in},{\mathrm{col}}}(t),b_{\mathrm{in},{\mathrm{col}}}^{\dag}(t')\right]=\delta(t-t')$. We then obtain the same quantum Langevin equations as for the single detector (\ref{ode}), but with the coupling scaled by $\sqrt{N}$:
\begin{equation}
\frac{d}{dt}
\begin{pmatrix}
a \\ a^{\dagger} \\ b_{\mathrm{col}} \\ b_{\mathrm{col}}^{\dagger}\\
\end{pmatrix}
=
\begin{pmatrix}
-\frac{\gamma}{2} &0 &0 &i\sqrt{N}\lambda\\
0 &-\frac{\gamma}{2} &-i\sqrt{N}\lambda &0\\
0 &i\sqrt{N}\lambda &-\frac{\gamma}{2} &0\\
-i\sqrt{N}\lambda &0 &0 &-\frac{\gamma}{2} \\
\end{pmatrix}
\begin{pmatrix}
a \\ a^{\dagger} \\ b_{\mathrm{col}} \\ b_{\mathrm{col}}^{\dagger}\\
\end{pmatrix}
+\sqrt{\gamma}
\begin{pmatrix}
a_{\mathrm{in}} \\ a_{\mathrm{in}}^{\dagger} \\ b_{\mathrm{in},\mathrm{col}} \\b_{\mathrm{in},\mathrm{col}}^{\dagger}\\
\end{pmatrix},
\label{ode2}
\end{equation}
where the phases $\phi_n$ drop out when performing the RWA. From Eq. (\ref{secondmoments1}), we therefore have that the steady state cavity mode average photon number with $N$ oscillating detectors  is
\begin{equation}
\left.\langle a^{\dag}(t) a(t)\rangle\right|_{t\rightarrow\infty}=\frac{2N \eta^2}{1-4N\eta^2},
\end{equation}
where $\eta=\lambda/\gamma$ here still refers to the single detector coupling to damping ratio.
The parametric instability threshold is then lowered to $\eta_{\mathrm{crit}}=1/(2 N)$. For $\eta$ well below the instability threshold, we see that the steady state cavity mode average photon number is scaled by $N$ compared to the single detector case. In essence, the oscillating detectors act incoherently in generating photons from the cavity mode ground state. 

If the detectors are modeled more appropriately by quantized nonlinear oscillators, (e.g., by two level systems), then we expect the instability threshold to be replaced instead by a quantum critical point; beyond the critical point, the detectors may instead act coherently with the steady state average cavity mode photon number now scaling as $N^2$, a manifestation of a non-equilibrium superradiant phase transition\cite{Bastidas,Wang}, thus significantly enhancing the Unruh effect for $N\gg 1$.

\section{\label{analogue}The FBAR-Superconducting Circuit Analogue}

Microwave superconducting circuits involving nonlinear Josephson junction (JJ) elements have proven a fruitful arena for investigating various photon production from vacuum analogues \cite{nation2012}, culminating in the experimental demonstration of an analogue of the dynamical Casimir effect (DCE) \cite{wilson2011}. In the latter analogue, the accelerating, oscillating mirror boundary of the electromagnetic vacuum is replaced by a flux tunable dc-SQUID at one end of a co-planar microwave cavity. By applying a sinusoidal, time varying magnetic flux through the SQUID, the effective length of the cavity that determines the microwave modes is modulated, resulting in photon pair production from vacuum under the right frequency conditions. 

However, a more satisfying demonstration of the DCE would involve an actual {\it moving}, i.e., mechanically oscillating mirror, rather than an electronic analogue. It was recently shown \cite{sanz2017} that by incorporating a capacitance at the end of the cavity with a few hundred nanometer thick dielectric layer undergoing dilatational oscillations at suitable GHz frequencies, potentially measurable photon production rates are predicted. Such mechanical resonators are commonly termed ``film bulk acoustic resonators" (FBARs) when piezoelectrically actuated \cite{oconnell2010}. We will adopt the same FBAR acronym to describe also thin, non-piezoelectric membranes undergoing dilatational motion. 

\begin{figure}
\begin{center}
\includegraphics[width=5in]{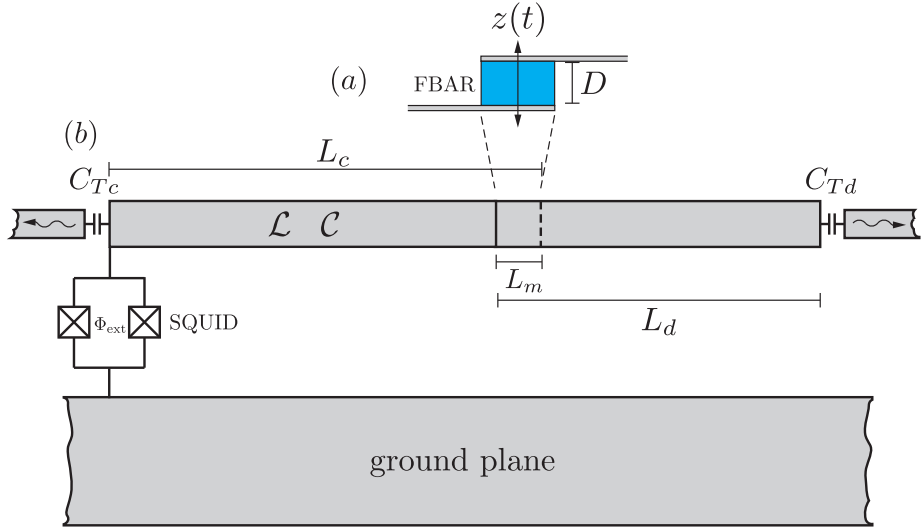} 
\caption{\label{schemefig} Scheme of the device. (a) Side view showing the FBAR resonator comprising a dielectric crystalline material with thickness $D$  sandwiched between the detector (upper) and cavity (lower) center conductors.  The FBAR  undergoes fundamental dilational mode displacements $z(t)$, modulating the distance between the cavity and detector center conductors. The detector center conductor is suspended for a segment of its length, in order to enable flexural mode coupling to the FBAR.  (b) Top view showing the cavity and detector resonators comprising their center conductors with lengths $L_c$ and $L_d$, respecively, and ground plane. The center conductors couple to each other via the FBAR capacitance of length $L_m$. Note that FBAR capacitance length is not drawn to scale: $L_m\ll L_c,\, L_d$.   The cavity and detector resonators are coupled to separate transmission lines via capacitances $C_{Tc}$ and $C_{Td}$, respectively, which feed to subsequent amplification stages for cross-correlation measurements of the resulting photon pair production. The flux-biased DC-SQUID enables tuning of the cavity mode frequency, so as to bring the sum of the two cavity mode frequencies into resonance with the FBAR mechanical drive frequency (parametric resonance condition). }
\end{center}
\end{figure}

In Fig. \ref{schemefig}, we show a possible practicable scheme inspired by the DCE proposal of Ref. \cite{sanz2017} that furnishes an Unruh effect analogue involving a mechanically oscillating detector based on an FBAR undergoing dilatational motion. The scheme comprises two coplanar microwave cavities with center conductors modeled as 1D strips having capacitance per unit length ${\mathcal{C}}$  and inductance per unit length ${\mathcal{L}}$. We denote one of the microwave resonators with center conductor length $L_c$ the ``cavity" (subscript `$c$'), while the other with center conductor length $L_d$ is denoted as the photon ``detector" (subscript `$d$'), although there is no particular distinction between cavity and detector  resonators for our scheme given that we model the photon detector as a harmonic oscillator as opposed to a two level system. The cavity and detector center conductors are deliberately chosen to have different lengths so that the resulting normal mode frequencies are sufficiently distinct (see discussion later below), with the center conductors overlapping at one end for a length $L_{m}$ to form the FBAR-capacitive coupling.  At the opposite end of the cavity center conductor is a dc-magnetic flux biased SQUID that serves to fine-tune its resonance frequency in order to appropriately match the relevant cavity and detector mode frequencies with the FBAR frequency as required for resonant photon pair production. 

The FBAR has a thickness $D$, which together with the added metallic layers of the center conductor on both sides forming the capacitive coupling,  determines the fundamental dilational mode frequency. In particular,  the FBAR is assumed to be driven with some steady state mechanical displacement $z(t)=A\cos\left(\Omega_m t\right)$, $A\ll D$, where $\Omega_{m}$ denotes the FBAR dilatational mode frequency and $A$ is the dilatational displacement amplitude. The subscript `$m$' here stands for ``mechanical", to emphasize that the parametric drive involves a mechanically oscillating degree of freedom, in this case the capacitor thickness that couples the cavity and photon detector modes. However, unlike the actual Unruh effect as described in the above sections, the present analogue does not involve the detector accelerating {\it within} the volume occupied by the cavity mode. 

In contrast to the DCE scheme of Ref. \cite{sanz2017}, we envisage utilizing a non-piezoelectric dielectric such as silicon between the overlapping center conductor capacitor plates for the FBAR; this avoids correlated photon pair production arising from induced, oscillating piezoelectric surface charges on the capacitor plates. Note however that in an actual device, care must be taken to address possible photon production due to spurious charges on the capacitor plate surfaces (i.e., electrostatic patch potentials). One possible way to actuate the dilational mode of the FBAR is via coupled $\sim 10~{\mathrm{GHz}}$ flexural vibrations in a suspended segment of the detector center conductor that are induced through a separate piezoelectric transducer located some distance away from the overlapping center conductor capacitor plates.  

\subsection{The analogue circuit cavity mode-detector Hamiltonian}
\label{detectsec}

The cavity and detector center conductors are  assumed to be weakly coupled capacitively to separate transmission lines that feed into measurement circuitry for verifying correlated photon production in the cavity-detector system. In the first part of the analysis, we neglect the capacitive coupling to the measurement circuitry and treat the cavity-detector system with driven FBAR as a closed system, deriving the mode Hamiltonian of the latter. We also neglect the SQUID element in the analysis, since it serves effectively as a flux dependent frequency ``tuner" and does not play an essential role in the dynamics. Applying Kirchhoff's laws to the circuit in Fig. \ref{schemefig} and performing a normal mode analysis of the closed cavity-detector system dynamics (see Appendix \ref{modemodelsec} for the details of the derivation), we obtain the following Hamiltonian:
\begin{equation}
H=\sum_{n} \hbar\omega_{n} {a}^{\dag}_{n} {a}_{n} -\frac{A}{D}\cos\left(\Omega_m t\right)\sum_{n,n'} \hbar \lambda_{nn'} \sqrt{\omega_{n}\omega_{n'}} \bigl({a}_{n}-{a}^{\dag}_{n}\bigr)\bigl({a}_{n'}-{a}^{\dag}_{n'}\bigr),
\label{creatannihhameq}
\end{equation}
where the label $n$ denotes the normal mode and the 
dimensionless coupling between the normal modes is given by the following formula: 
\begin{equation}
\lambda_{n n'}=\left(\frac{\pi}{\Phi_0}\right)^2 \frac{{\mathcal{C}}_{m}}{\sqrt{C_n C_{n'}}}\int_{0}^{L_{m}} dx \left[\Phi_{d,n}(L_d-x)-\Phi_{c,n}(L_c-x)\right]\left[\Phi_{d,n'}(L_d-x)-\Phi_{c,n'}(L_c-x)\right].
\label{couplingeq}
\end{equation}
Here, $\Phi_{d, n } (x)$ and $\Phi_{c, n }(x)$ are the normal mode flux field solutions in the detector and cavity resonator, while $\Phi_0=h/(2 e)$ is the flux quantum. The parameter ${\mathcal{C}}_{m}$ denotes the undisplaced FBAR capacitance per unit center conductor length, which we assume to be well approximated by the parallel plate capacitance formula; the $C_n$'s are mode normalization constants with the dimensions of capacitance. 

Note that  for the strong capacitive couplings we consider between the cavity and detector resonators (see below), the modes of each of these subsystems become strongly hybridized, with the resulting normal modes having non-negligible amplitude in both the cavity and detector resonator. The operators $a_n^{\dag}$ thus create photons that coexist in the two resonator regions and their physical identification as cavity and detector is no longer that meaningful. Instead, it is more appropriate to relate the ``cavity" and ``detector" labels (or ``idler" and ``signal") in the analogue to the respective normal modes $n=1,\, 2$.

Tuning the frequencies such that $\Omega_{m}=\omega_1+\omega_2$, we can simplify the Hamiltonian (\ref{creatannihhameq}) by transforming to the interaction picture and making a rotating wave approximation to obtain
\begin{equation}
{H}_I=\hbar\lambda \left({a}_{1}{a}_{2}+{a}^{\dag}_{1}{a}^{\dag}_{2}\right),
\label{rwadpeheq}
\end{equation}
where we define here $\lambda=-\lambda_{12}\sqrt{\omega_{1}\omega_2}A/D$; recall that $A$ is the mechanical, dilatational displacement amplitude and $D$ is the FBAR thickness. This Hamiltonian coincides with the standard, non-degenerate parametric amplifier (NDPA) Hamiltonian in the interaction picture.

We now estimate how large the coupling $\lambda$ can be, given realistic device parameters.  Assuming, e.g.,  cavity and detector normal mode frequencies $\omega_1\sim 2\pi\times 4\, {\mathrm{GHz}}$ and $\omega_2\sim 2\pi\times 6\, {\mathrm{GHz}}$ , we require the FBAR dilatational frequency 
to be $\Omega_{m}\sim 2\pi\times 10~{\mathrm{GHz}}$. Approximating the FBAR as elastically isotropic and neglecting the mechanical contribution from the metal plates (we can assume that their thicknesses are much less than $D$), the fundamental dilatational mode frequency of the FBAR is given by the expression \cite{oconnell2010}
\begin{equation}
\Omega_{m}=\frac{\pi v_l}{D},
\label{dilatfreq}
\end{equation}
where $v_l$ is the propagation speed of a longitudinal elastic plane wave. Assuming silicon for the FBAR material, we have $v_l\approx 10^4~ {\mathrm{m}}/{\mathrm{s}}$, and from the required $\Omega_{{m}}$ frequency, the FBAR thickness must be approximately $D\approx 500~{\mathrm{nm}}$. Using the parallel plate capacitance expression for the overlapping center conductor plates with gap thickness $D=500~{\mathrm{nm}}$ and dielectric constant $\varepsilon_r\approx 12$ for silicon, we obtain ${\mathcal{C}}_{m}\approx2\times 10^{-9}~{\mathrm{F}}/{\mathrm{m}}$ for a $10~\mu{\mathrm{m}}$ width (coinciding with the assumed center conductor width). For the cavity and detector resonators, we assume the capacitance and inductance per unit length to be ${\mathcal{C}}=  10^{-10}~{\mathrm{F}}/{\mathrm{m}}$ (about twenty times smaller than ${\mathcal{C}}_{m}$)  and ${\mathcal{L}}=  10^{-8}~{\mathrm{H}}/{\mathrm{m}}$, respectively.  

Consider example cavity and detector conductor lengths $L_c =1.1~{\mathrm{cm}}$ and $L_d=0.8~{\mathrm{cm}}$. Through the expression $\omega=\pi v/L$ for the half-wave  mode of a single, uncoupled cavity resonator, with electromagnetic wave speed in the cavity given by $v=1/\sqrt{{\mathcal{L}}{\mathcal{C}}}$, these lengths correspond to frequencies $\omega_1/(2\pi)=4.5~{\mathrm{GHz}}$ and $\omega_2/(2\pi)=6.5~{\mathrm{GHz}}$. While these are larger than our desired frequencies, note that introducing the FBAR coupling capacitance ${\mathcal{C}}_{m} L_{m}$ will result in lower normal mode frequencies due to the increase in capacitance.

\begin{figure}
\begin{center}
\includegraphics[width=5in]{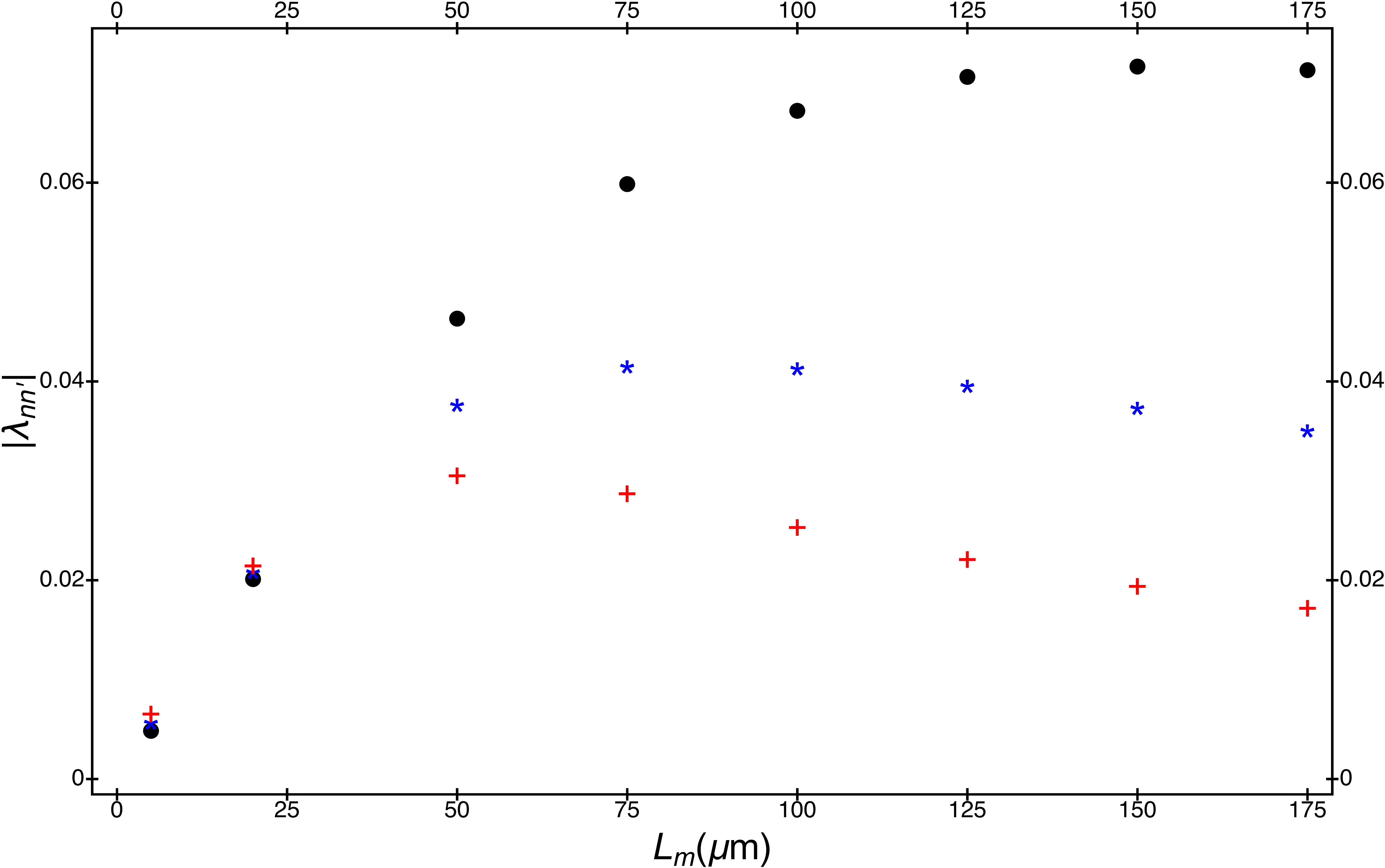} 
\caption{\label{couplingfig} Coupling strength $\lambda_{nn'}$ dependence on FBAR capacitance length $L_{m}$; `$\cdot$' denotes $\lambda_{11}$, `$*$' denotes $\lambda_{12}$, and `$+$' denotes $\lambda_{22}$.}
\end{center}
\end{figure}
\begin{figure*}[thb]
\centering
\subfloat[]{\label{m1fig}
  \includegraphics[width=0.45\textwidth]{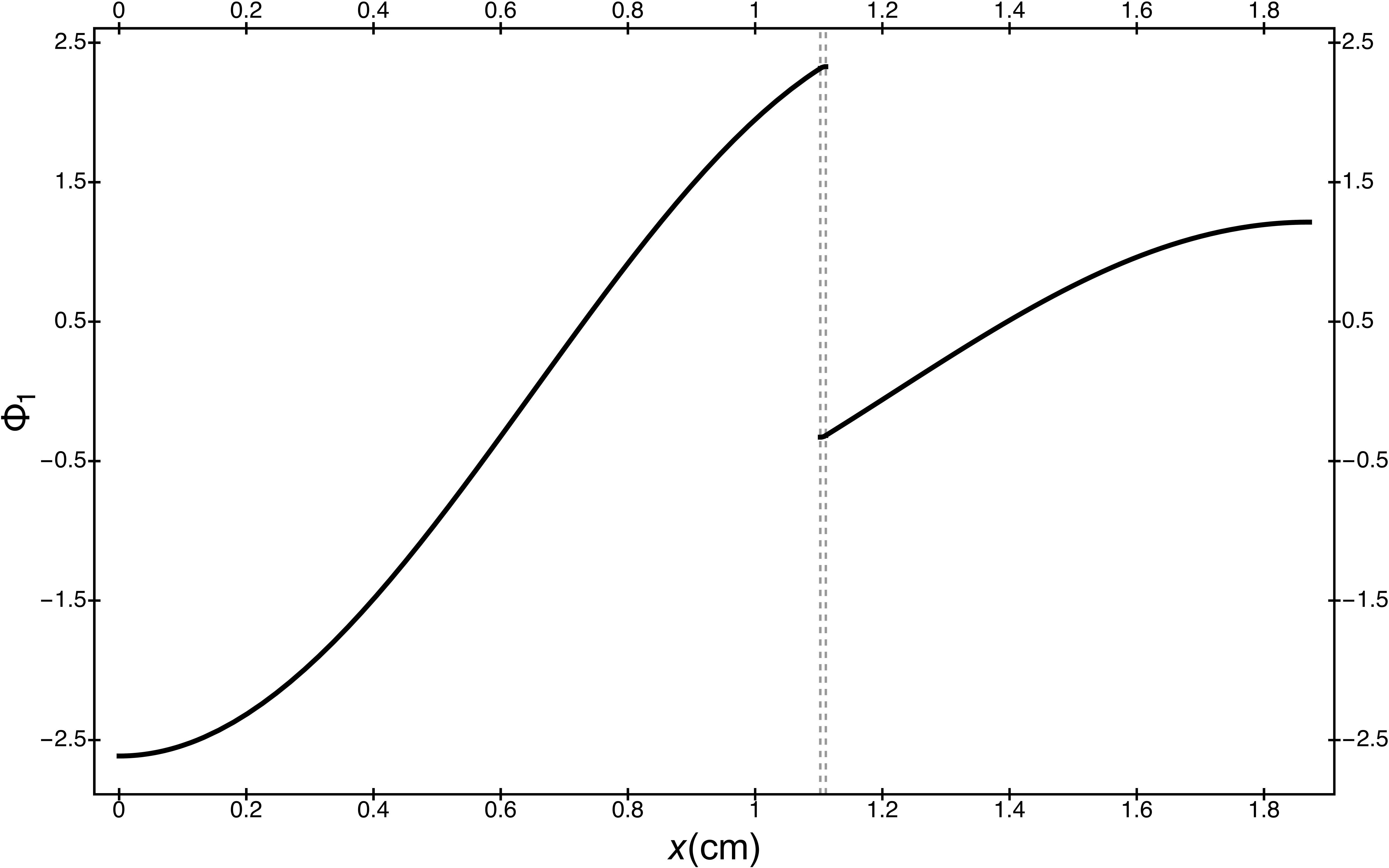}
}
\subfloat[]{\label{m2fig}
  \includegraphics[width=0.45\textwidth]{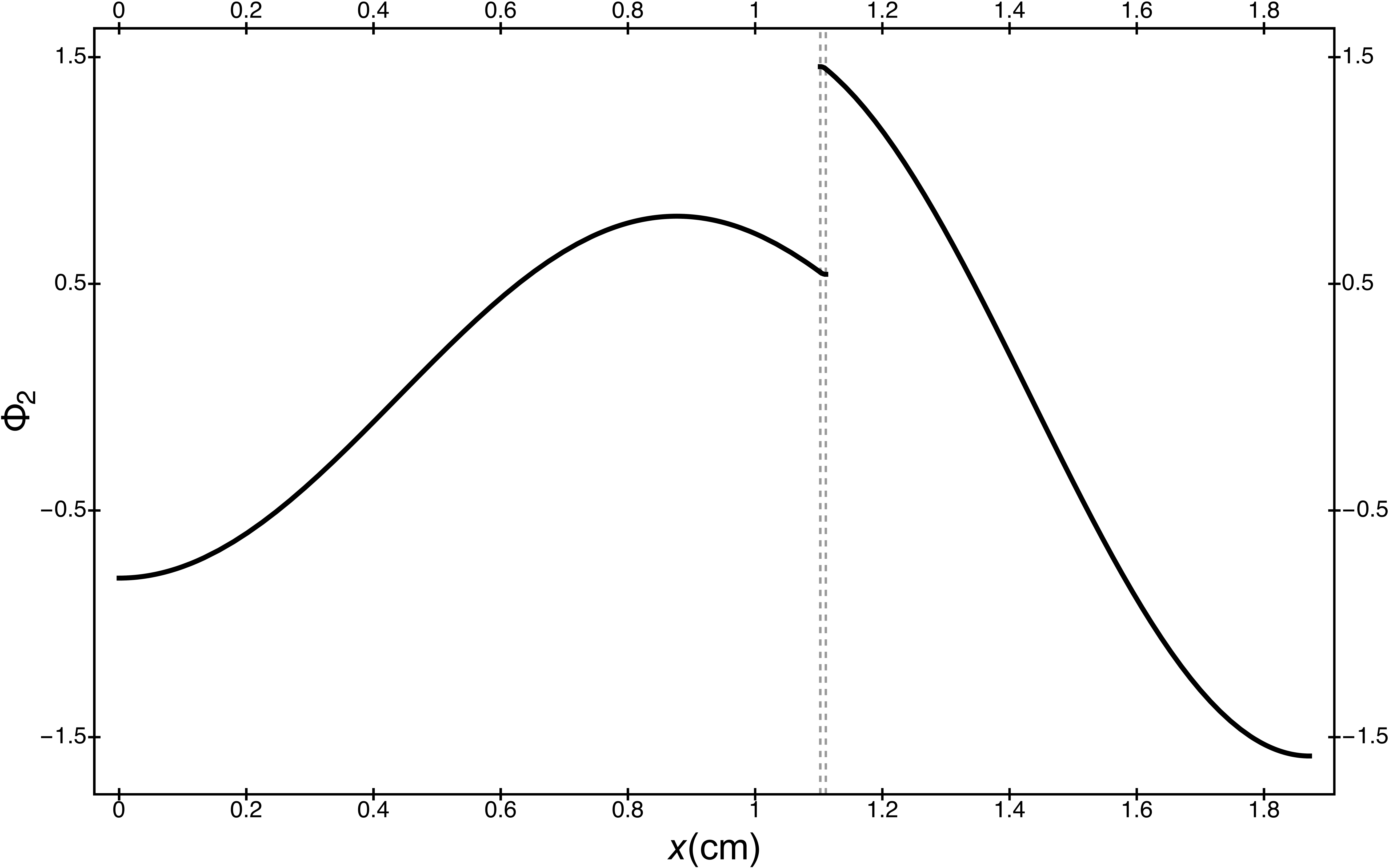}}
\caption{\label{modefig} Normal mode functions (arbitrary scale)  (a) $\Phi_1 (x)$ and  (b) $\Phi_2(x)$. Note that the $x$ coordinate is defined here such that $x=0$ corresponds to the left end of the cavity center conductor and $x=L_c+L_d-L_m$ corresponds to the right end of the detector center conductor (see Fig. \ref{schemefig}). The vertical dashed lines indicate the location of the FBAR coupling capacitor ($L_c-L_m\leq x\leq L_c$).  }
\end{figure*}

Evaluating the coupling strengths (\ref{couplingeq}) versus FBAR capacitance length $L_{m}$ for the lowest frequency modes $n=1,\, 2$, we obtain the results shown in Fig. \ref{couplingfig}. The coupling strengths scale approximately linearly with $L_{m}$ for small values, with the linear dependence breaking down when ${\mathcal{C}}_{m} L_{m}$ is within an order of magnitude of the center conductor capacitances ${\mathcal{C}}L_c$ and ${\mathcal{C}}L_d$, corresponding to $L_m\sim 40~\mu{\mathrm{m}}$.  Beyond this capacitance length, the cavity and detector modes become strongly hybridized; the coupling between normal modes $n=1$ and $n'=2$ has a maximum (in magnitude) $\left|\lambda_{12}\right|=0.04$ for $L_{m}=90~\mu{\mathrm{m}}$, corresponding to normal mode frequencies $\omega_1/(2 \pi)=3.8~{\mathrm{GHz}}$ and $\omega_2/(2 \pi)=5.7~{\mathrm{GHz}}$. Note that these frequencies are smaller than the original uncoupled cavity and detector frequencies $\omega_1/(2\pi)=4.5~{\mathrm{GHz}}$ and $\omega_2/(2\pi)=6.5~{\mathrm{GHz}}$, respectively, signifying the strong renormalizing effects of the coupling capacitance ${\mathcal{C}}_{m} L_{m}$. Fig. \ref{modefig} shows the normal mode solutions $\Phi_1(x)$ and $\Phi_2(x)$ for which the coupling $|\lambda_{12}|$ is a maximum. Note the strong hybridization of the cavity and detector modes, extending throughout the cavity and detector lengths. With the FBAR thickness $D=500~{\mathrm{nm}}$ and assuming an achievable dilatational amplitude $A=10^{-11}~{\mathrm{m}}$ \cite{cuffe2013}, we obtain the maximum coupling strength $\lambda\approx 24.5 \times 10^3~{\mathrm{Hz}}$  in Eq. (\ref{rwadpeheq}).

In the above derivation of the FBAR-superconducting circuit Hamiltonian, we treated the dilatational motion non-relativistically, i.e., we neglected Lorentz time dilation factors. This is well-justified for our analogue since with the assumed parameter values, the dilatational velocity magnitude of the FBAR surfaces is $A\Omega_m \sim 10^{-11}~{\mathrm{m}}\times 2\pi \times 10^{10}~{\mathrm{Hz}}\sim 1~ {\mathrm{m}}/{\mathrm{s}}\lll c$.    

\subsection{Measurement scheme}
\label{measurementsec}
In the following, we describe how the generated correlated photon pairs may be measured  that result from the nondegenerate parametric amplification of resonator mode vacuum fluctuations with closed system Hamiltonian given by Eq. (\ref{rwadpeheq}).  To proceed, we must take into account the (weak) capacitive couplings between the detector and cavity resonators and their respective  transmission lines (denoted by $C_{Td}$ and $C_{Tc}$, respectively in Fig. \ref{schemefig}) that feed into the subsequent signal amplification stages.

Including the weakly coupled transmission lines, the cavity and detector resonator modes will as a consequence be damped and subject to electromagnetic noise. The resulting quantum dynamics can be described to a good approximation using the ``input-output" approach \cite{gardiner1985,gardiner2000} by the following quantum Langevin equations: 
 \begin{eqnarray}
 \frac{d{a}_1}{dt}&=&\frac{i}{\hbar}\left[{H},{a}_1\right]-\frac{1}{2}\left(\gamma_{c 1} +\gamma_{d1}\right) {a}_1 +i\sqrt{\gamma_{c1}} {a}_{ c}^{\mathrm{in}}-i\sqrt{\gamma_{d1}} {a}_{ d}^{\mathrm{in}},\label{a1langeq}\\
  \frac{d{a}_2}{dt}&=&\frac{i}{\hbar}\left[{H},{a}_2\right]-\frac{1}{2}\left(\gamma_{c 2} +\gamma_{d1}\right) {a}_2 +i\sqrt{\gamma_{c2}} {a}_{ c}^{\mathrm{in}}+i\sqrt{\gamma_{d2}} {a}_{ d}^{\mathrm{in}},\label{a2langeq}
 \end{eqnarray}
 where the Hamiltonian (\ref{rwadpeheq}) in the Heisenberg picture is 
\begin{equation}
H=\hbar\omega_1 a_1^{\dag}a_1+\hbar\omega_2 a_2^{\dag}a_2 +\hbar\lambda \left(e^{-i\Omega_{m}t}a_1^{\dag}a_2^{\dag}+e^{i\Omega_{m}t}a_1 a_2\right),
\label{hameq}
\end{equation}
with $\Omega_m=\omega_1+\omega_2$.
In Eqs. (\ref{a1langeq}) and (\ref{a2langeq}), the $n=1(2)$ mode damping rates $\gamma_{c1(2)}$ result from the capacitive coupling of the cavity resonator to the left transmission line, while the damping rates $\gamma_{d1(2)}$ result from the capacitive coupling of the detector resonator to the right transmission line (see Fig.~\ref{schemefig}). The sign differences in the noise terms follow from the relative signs of the mode functions $\Phi_1$ and $\Phi_2$ at the cavity and detector center conductor ends that are coupled to their respective transmission lines. Equations (\ref{a1langeq}) and (\ref{a2langeq}) are accompanied by the input-output relations
\begin{eqnarray}
{a}_{ c}^{\mathrm{out}}- {a}_{ c}^{\mathrm{in}}&=&i \sqrt{\gamma_{c1}} {a}_1+i \sqrt{\gamma_{c2}} {a}_2,
 \label{iorel1eq}\\
 {a}_{ d}^{\mathrm{out}}- {a}_{ d}^{\mathrm{in}}&=&-i \sqrt{\gamma_{d1}} {a}_1 +i \sqrt{\gamma_{d2}} {a}_2.
 \label{iorel2eq}
 \end{eqnarray}
 
Taking the Fourier transform [i.e., $f(\omega)=(2\pi)^{-1/2}\int_{-\infty}^{+\infty} dt e^{i\omega t}f(t)$] of the quantum Langevin equations (\ref{a1langeq}) and (\ref{a2langeq}) and solving for ${a}_{1(2)}(\omega)$, we obtain
\begin{eqnarray}
{a}_1(\omega)&=&\left[\left(-i(\omega-\omega_1)+\frac{\gamma_1}{2}\right)\left(-i(\omega-\omega_1)+\frac{\gamma_2}{2}\right) -\lambda^2\right]^{-1}\cr
&\times&\left\{\left[i\sqrt{\gamma_{c1}}{a}^{\mathrm{in}}_{ c}(\omega)-i\sqrt{\gamma_{d1}}{a}^{\mathrm{in}}_{ d}(\omega) \right]\left[-i(\omega-\omega_1)+\frac{\gamma_2}{2}\right]\right.\cr
&-&\left.\lambda\left[\sqrt{\gamma_{c2}}\left({a}^{\mathrm{in}}_{ c}(\Omega_m-\omega)\right)^{\dag}+\sqrt{\gamma_{d2}}\left({a}^{\mathrm{in}}_{ d}(\Omega_m-\omega)\right)^{\dag} \right]
\right\},
\label{a1solneq}
\end{eqnarray}
\begin{eqnarray}
{a}_2(\omega)&=&\left[\left(-i(\omega-\omega_2)+\frac{\gamma_1}{2}\right)\left(-i(\omega-\omega_2)+\frac{\gamma_2}{2}\right) -\lambda^2\right]^{-1}\cr
&\times&\left\{\left[i\sqrt{\gamma_{c2}}{a}^{\mathrm{in}}_{ c}(\omega)+i\sqrt{\gamma_{d2}}{a}^{\mathrm{in}}_{ d}(\omega) \right]\left[-i(\omega-\omega_2)+\frac{\gamma_1}{2}\right]\right.\cr
&-&\left.\lambda\left[\sqrt{\gamma_{c1}}\left({a}^{\mathrm{in}}_{ c}(\Omega_m-\omega)\right)^{\dag}-\sqrt{\gamma_{d1}}\left({a}^{\mathrm{in}}_{ d}(\Omega_m-\omega)\right)^{\dag} \right]
\right\},
\label{a2solneq}
\end{eqnarray}
where $\gamma_1=\gamma_{c1}+\gamma_{d1}$ and $\gamma_2=\gamma_{c2}+\gamma_{d2}$. 
Substituting Eqs. (\ref{a1solneq}) and (\ref{a2solneq}) into the Fourier transform of the input-output relations (\ref{iorel1eq}) and (\ref{iorel2eq}), we obtain solutions to ${a}^{\mathrm{out}}_{ c}(\omega)$ and ${a}^{\mathrm{out}}_{ d}(\omega)$, which can then be used to calculate various measurable quantities. The latter are  expressed in terms of the filtered output, transmission line voltage or current operators:
\begin{equation}
{V}^{\mathrm{out}}_{c(d)}(t)=-i\left(\frac{\hbar Z_T}{4\pi}\right)^{1/2}\int_{\omega_0-\Delta\omega/2}^{\omega_0+\Delta\omega/2}d\omega \sqrt{\omega}\left[e^{-i\omega t} a^{\mathrm{out}}_{c(d)} \left(\omega\right)-e^{i\omega t} \left(a^{\mathrm{out}}_{c(d)} \left(\omega\right)\right)^{\dag}\right] 
\label{vouteq}
\end{equation}
and 
\begin{equation}
{I}^{\mathrm{out}}_{c(d)}(t)=i\left(\frac{\hbar}{4\pi  Z_T}\right)^{1/2}\int_{\omega_0-\Delta\omega/2}^{\omega_0+\Delta\omega/2}d\omega \sqrt{\omega}\left[e^{-i\omega t} a^{\mathrm{out}}_{c(d)} \left(\omega\right)-e^{i\omega t} \left(a^{\mathrm{out}}_{c(d)} \left(\omega\right)\right)^{\dag}\right], 
\label{iouteq}
\end{equation}
where the filter bandwidth $\Delta \omega$ is centered at frequency $\omega_0$, and $Z_T$ is the impedance of the transmission lines.

We first determine the filtered output power in some bandwidth $\Delta\omega$ centered at  frequency $\omega_0$; as we shall see, the output power gives a measure of the rate at which photons are produced from vacuum by the FBAR mechanical parametric drive. The power radiated into the  transmission line connected to the cavity resonator is
\begin{equation}
P_{ c}^{\mathrm{out}}=\left\langle I_c^{\mathrm{out}\, 2 }\right\rangle Z_T=\frac{\hbar}{4\pi}\int_{\omega_0-\Delta\omega/2}^{\omega_0+\Delta\omega/2}d\omega d\omega' \omega\left[\langle a_{ c}^{\mathrm{out}}(\omega) \left(a_{ c}^{\mathrm{out}}(\omega')\right)^{\dag}\rangle+\langle \left(a_{ c}^{\mathrm{out}}(\omega)\right)^{\dag} a_{ c}^{\mathrm{out}}(\omega')\rangle\right],
\label{outpowereq}
\end{equation}
where the angular brackets denote an ensemble average with respect to the `in' states of the transmission line, and we have also performed a time average; a similar expression also holds for the power radiated into the  transmission line connected to the detector resonator ($c\leftrightarrow d$). Given that each of the two normal mode functions extend throughout both coupled cavities, we effectively have a so-called `two-sided' cavity \cite{dasilva2010}. This affords the alternative possibility of a cross-correlated measurement of the currents in both transmission lines to determine the radiated power:
\begin{eqnarray}
P_{ c d}^{\mathrm{out}}&=&\frac{1}{2}\left\langle \left(I_c^{\mathrm{out}} I_d^{\mathrm{out}}+I_d^{\mathrm{out}} I_c^{\mathrm{out}}\right)\right\rangle Z_T\cr
&=&\frac{\hbar}{8\pi}\int_{\omega_0-\Delta\omega/2}^{\omega_0+\Delta\omega/2}d\omega d\omega' \omega\left[\langle a_{ c}^{\mathrm{out}}(\omega) \left(a_{ d}^{\mathrm{out}}(\omega')\right)^{\dag}\rangle+\langle \left(a_{ c}^{\mathrm{out}}(\omega)\right)^{\dag} a_{ d}^{\mathrm{out}}(\omega')\rangle +{\mathrm{h.c.}}\right],
\label{outpower2eq}
\end{eqnarray}
where `${\mathrm{h.c.}}$' denotes ``Hermitian conjugate".
The advantage of considering the cross correlation between the current outputs of the two transmission lines over the autocorrelation between the current outputs of a single transmission line, is that additive transmission line noise does not arise in the output signal.  This is because the cavity-connected transmission line noise operator commutes with the detector-connected transmission line noise operator: $[{a}^{\mathrm{in}}_c (\omega),\left({a}^{\mathrm{in}}_d (\omega')\right)^{\dag}]=0$. Most crucially, taking into account the necessary, subsequent amplification of the output signals in each transmission line and then cross correlating, the added noise of the amplifiers would be significantly reduced, limited only by possible weak correlations between the cavity and detector amplifier added noise modes \cite{dasilva2010}. 

Substituting the solutions (\ref{a1solneq}) and (\ref{a2solneq}) into Eq. (\ref{outpower2eq}), we obtain for the cross-correlated power in bandwidth $\Delta\omega$ about frequency $\omega_0$:
\begin{eqnarray}
&&P_{cd}^{\mathrm{out}}=\int_{\omega_0-\Delta\omega/2}^{\omega_0+\Delta\omega/2}\frac{d\omega}{2\pi} \hbar\omega (2\bar{n}+1)\lambda^2\cr
&&\times\left\{-\sqrt{\gamma_{c1}\gamma_{d1}}\gamma_2\left|\left(-i(\omega-\omega_1)+\frac{\gamma_1}{2}\right)\left(-i(\omega-\omega_1)+\frac{\gamma_2}{2}\right) -\lambda^2\right|^{-2}\right.\cr 
&&\left.+\sqrt{\gamma_{c2}\gamma_{d2}}\gamma_1\left|\left(-i(\omega-\omega_2)+\frac{\gamma_1}{2}\right)\left(-i(\omega-\omega_2)+\frac{\gamma_2}{2}\right) -\lambda^2\right|^{-2}\right\},
\label{outpower3eq}
\end{eqnarray}
where $\bar{n}=(e^{\hbar\omega/(k_BT)}-1)^{-1}$ is the thermal average photon occupation number of the transmission lines (we assume both transmission lines are at the same temperature $T$), and we have also assumed $\gamma_{cn},\, \gamma_{dn}\ll\omega_n,\, n=1,2$. The correlated power comprises the sum of two contributions, corresponding to the parametric amplification of thermal and zeropoint fluctuations in modes 1 and 2. In particular, in the limit of vanishing temperature such that $\bar{n}=0$, the remaining non-zero contribution to the output power arises from the pair-production of photons in modes 1 and 2 out of their vacuum (i.e., ground) state. Note that the minus sign in front of the mode 1 term is due to the opposite signs of the associated mode function at the outer ends of the coupled cavities (Fig. \ref{modefig}).

\begin{figure}
\begin{center}
\includegraphics[width=5in]{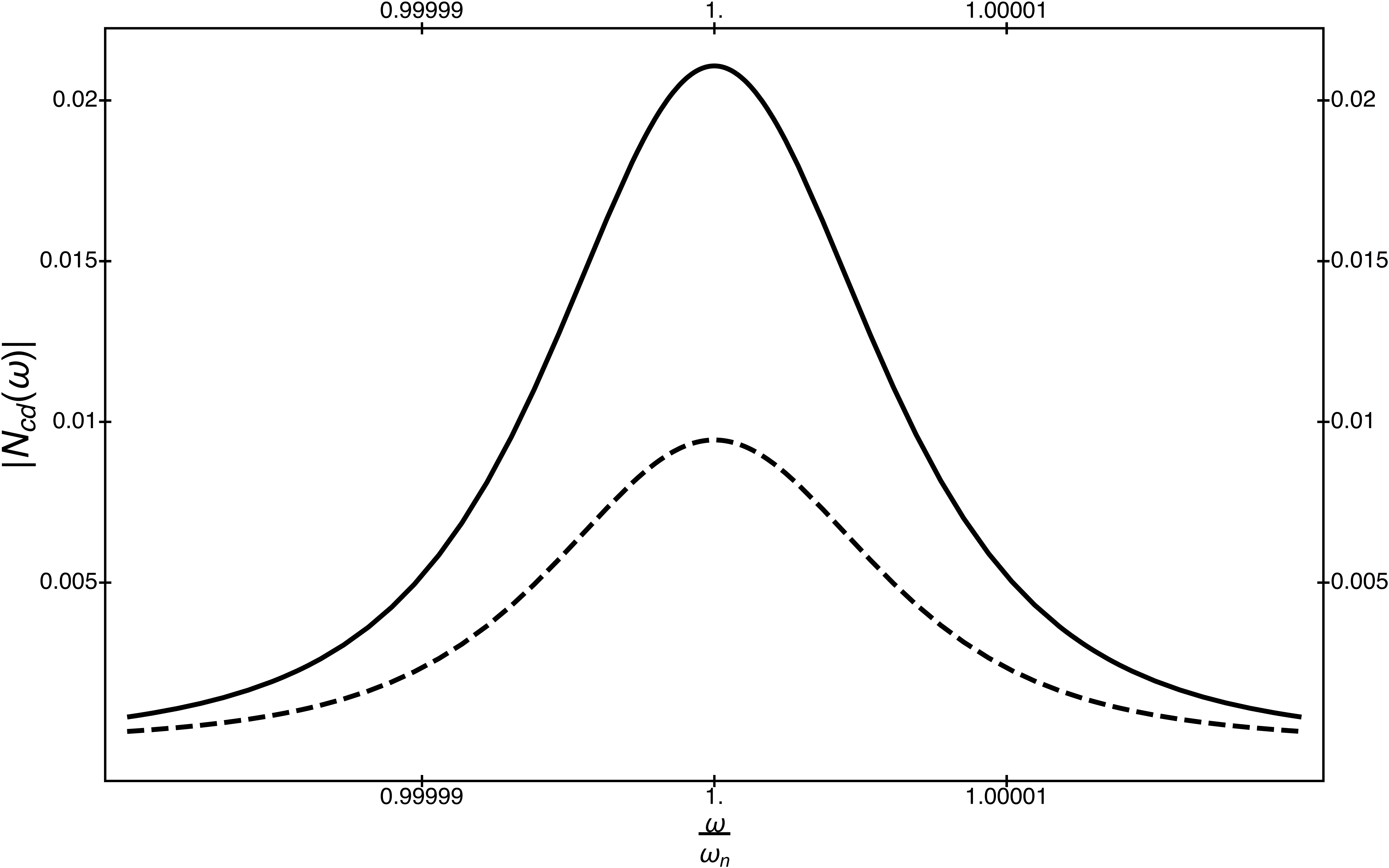} 
\caption{\label{nplotfig} Cross-correlated  vacuum photon emission rate per unit Hertz ${{N}}_{ c d}(\omega)$ versus frequency about mode $n=1$ (solid line) and mode $n=2$ (dashed line).}
\end{center}
\end{figure}

From Eq. (\ref{outpower3eq}), we define the cross-correlated power spectral density $S_{c d}(\omega)$ through the relation
\begin{equation}
P^{\mathrm{out}}_{c d}=\int_{\omega_0-\Delta/2}^{\omega_0+\Delta/2}\frac{d\omega}{2\pi} S_{c d}(\omega) 
\label{powerdenseq}
\end{equation}
and also define the cross-correlated photon emission rate per unit Hertz as ${{N}}_{c d}(\omega)=S_{c d}(\omega)/(\hbar\omega)$. Figure \ref{nplotfig}  shows a plot of the vacuum ($\bar{n}=0$), cross-correlated photon emission rate  ${{N}}_{ c d}(\omega)$ per unit Hertz versus frequency $\omega$ for the sample realisable parameter values considered above in Sec. \ref{measurementsec}, i.e., $\omega_1/(2 \pi)=3.8~{\mathrm{GHz}}$, $\omega_2/(2 \pi)=5.7~{\mathrm{GHz}}$, and $\lambda\approx 24.5 \times 10^3~{\mathrm{Hz}}$. We also assume identical, realizable cavity and detector resonator cavity quality factors $Q_1=Q_2=10^5$, so that the cavity damping rates ($\gamma_n=\omega_n/Q$) are $\gamma_{1}=2.4\times 10^5~{\mathrm{s}}^{-1}$ and $\gamma_{2}=3.6\times 10^5~{\mathrm{s}}^{-1}$. With such damping rates, we have $\lambda/\gamma_1=0.1$ and $\lambda/\gamma_2=0.07$, which are about an order of magnitude below the parametric instability onset; the photon emission rate maximum scales approximately linearly with inverse damping rate and quadratically with dilatational mode displacement amplitude when well below this instability onset. For the above parameter values, the peak cross-correlated photon emission rate per Hertz is $|{{N}}_{ c d}|\approx 0.02$ at $\omega=\omega_1$ and $|{{N}}_{ c d}|\approx 0.01$ at $\omega=\omega_2$ (see Fig. \ref{nplotfig}). These correspond to peak power densities $|S_{ c d}|/k_B\approx 4~{\mathrm{mK}}$ and $3 ~{\mathrm{mK}}$, respectively.
These signal levels are within reach of state-of-the-art cryogenic microwave techniques. In Ref. \cite{chang2018}, for instance, the covariance between modes of a parametric cavity was measured to milliKelvin precision using a conventional HEMT amplifier.  The averaging times in that work were of order three hours, suggesting that the precision could be further improved simply by averaging longer. If instead the HEMT amplifier was replaced by a quantum-limited Josephson parametric amplifier, the averaging times could be drastically reduced.

With just an order of magnitude reduction in the resonator damping rates from those assumed above, the power spectral density could be considerably enhanced by adjusting the coupling $\lambda$ through for example  fine tuning the mechanical drive amplitude $A\sim 10^{-11}~{\mathrm{m}}$ so as to remain just below the instability onset (c.f. Fig. \ref{tempentanglefig}).

While the cross-correlated power gives a measure of the parametric photon production rate without the presence of transmission line noise, it cannot distinguish photons created out of vacuum zero-point fluctuations (Unruh effect analogue) from the parametric amplification of thermal radiation ($\bar{n}\neq 0$) that might be the result of heating due to the mechanical FBAR actuation process. Quantum correlated photon pair production can be verified through for example quantum squeezing (so-called ``two-mode" squeezing) or quantum entanglement measures (e.g., logarithmic negativity--see Sec. \ref{sec:entanglement})  of the $n=1,\, 2$ cavity-detector resonator normal modes \cite{johansson2013}. In the remainder of this section, we analyze the two-mode squeezing and briefly discuss how it may be measured.

For two-mode squeezing, the appropriate observables to consider are the following superposition quadrature operators \cite{gerry2005}:
\begin{equation}
{X}_{1}(t)=2^{-3/2}\left[e^{i\omega_{1} t} {a}_1(t)+e^{-i\omega_{1} t} {a}^{\dag}_1(t)+e^{i\left(\omega_2 t-\theta\right)}{a}_2(t)+e^{-i\left(\omega_2 t-\theta\right)}{a}^{\dag}_2(t)\right]
\label{quadphase1eq}
\end{equation}
and
\begin{equation}
{X}_{2}(t)=-2^{-3/2}i\left[e^{i\omega_{1} t} {a}_1(t)-e^{-i\omega_{1} t} {a}^{\dag}_1(t)+e^{i\left(\omega_2 t-\theta\right)}{a}_2(t)-e^{-i\left(\omega_2 t-\theta\right)}{a}^{\dag}_2(t)\right].
\label{quadphase2eq}
\end{equation}
These two quadrature operators are complementary, satisfying the commutation relation
\begin{equation}
\left[{X}_{1}(t),{X}_{2}(t)\right]=\frac{i}{2}
\label{quadcreq}
\end{equation}
and the Heisenberg uncertainty principle 
\begin{equation}
\Delta X_{1}\Delta X_{2}\geq\frac{1}{4}.
\label{phasehupeq}
\end{equation}
For coherent states, we have $\Delta X_{1}=\Delta X_{2}=1/2$; we therefore define a ``quantum squeezed'' state as one for which $\min_{\theta}\Delta X_{1}<1/2$, i.e., where the phase angle $\theta$ is chosen such that $\Delta X_{1}$ is a minimum. 

In order to evaluate $\Delta X_{1}$ for the cavity-detector modes in the parametrically driven steady state, it is convenient to express the solution in terms of the Fourier transform of the quadrature operator:
\begin{equation}
\Delta X_{1}^2=\langle X_{1}(t)^2\rangle=\frac{1}{2\pi}\iint d\omega d\omega' e^{-i (\omega-\omega')t}\langle X_{1}(\omega)X_{1}(-\omega')\rangle,
\label{ftquad2eq}
\end{equation}
where
\begin{eqnarray}
{X}_{1}(\omega)&=&\frac{1}{\sqrt{2\pi}}\int_{-\infty}^{+\infty}dt e^{i\omega t}{X}_{1}(t)\cr 
&=&2^{-3/2}\left[a_1(\omega_1+\omega)+\left(a_1(\omega_1-\omega)\right)^{\dag}+a_2(\omega_2+\omega)e^{-i\theta}+\left(a_2(\omega_2-\omega)\right)^{\dag}e^{i\theta}\right].
\label{quadfteq}
\end{eqnarray}
From Eqs. (\ref{a1solneq}), (\ref{a2solneq}), and (\ref{quadfteq}), we obtain
\begin{eqnarray}
&&\langle X_{1}(\omega)X_{1}(-\omega')\rangle =\frac{1}{4}\delta(\omega-\omega')\left|\left(-i\omega+\frac{\gamma_1}{2}\right)\left(-i\omega+\frac{\gamma_2}{2}\right) -\lambda^2\right|^{-2}\cr &\times&\biggl\{-\left(\bar{n}_1+\bar{n}_2+1\right)\lambda\gamma_1\gamma_2\sin\theta-\lambda\left[\gamma_1(2\bar{n}_1+1)-\gamma_2(2\bar{n}_2+1)\right]\omega\cos\theta\cr 
&&+(2\bar{n}_1+1)\frac{\gamma_1}{2}\left[\omega^2+\frac{\gamma_2^2}{4} +\lambda^2\right]+(2\bar{n}_2+1)\frac{\gamma_2}{2}\left[\omega^2+\frac{\gamma_1^2}{4} +\lambda^2\right]\biggr\},
\label{ftxxeq}
\end{eqnarray}
where $\bar{n}_{1(2)}=(e^{\hbar\omega_{1(2)}/(k_BT)}-1)^{-1}$ (we assume that $\gamma_1\ll\omega_1$, $\gamma_2\ll\omega_2$). 
Substituting Expression (\ref{ftxxeq}) into (\ref{ftquad2eq}) and carrying out the angular frequency integrals, 
\begin{equation}
\Delta X_{1}^2=\frac{1}{4}(\bar{n}_1+\bar{n}_2+1)\frac{1-\frac{2\lambda}{(\gamma_1+\gamma_2)}\sin\theta}{1-\frac{\lambda^2}{\gamma_1\gamma_2}}.
\label{quadsqrdeq}
\end{equation}
This expectation value is a minimum for $\theta=\pi/2$ (with $\lambda>0$):
\begin{equation}
\Delta X_{1}^2=\frac{1}{4}(\bar{n}_1+\bar{n}_2+1)\left(\frac{1-\frac{2\lambda}{\gamma_1+\gamma_2}}{1-\frac{\lambda^2}{\gamma_1\gamma_2}}\right),
\label{quadmineq}
\end{equation}
with the complementary quadrature variance given by
\begin{equation}
\Delta X_{2}^2=\frac{1}{4}( \bar{n}_1+\bar{n}_2+1)\left(\frac{1+\frac{2\lambda}{\gamma_1+\gamma_2}}{1-\frac{\lambda^2}{\gamma_1\gamma_2}}\right).
\label{quad2mineq}
\end{equation}
The condition $\Delta X_{1}<1/2$ to be in the quantum squeezing regime then becomes
\begin{equation}
\bar{n}_1+\bar{n}_2<\left(\frac{1-\frac{\lambda^2}{\gamma_1\gamma_2}}{1-\frac{2\lambda}{\gamma_1+\gamma_2}}\right)-1.
\label{quantsqueezeq}
\end{equation}
For the above example parameter values, we require $\bar{n}_1+\bar{n}_2< 0.08$, which gives for the cavity-detector environment (i.e., transmission line) temperature $T<70~{\mathrm{mK}}$ in order to be in the quantum squeezing regime. Verifying this two-mode quantum squeezing requires measuring cross-correlations between the quadratures of the output transmission line voltage (\ref{vouteq}) or current (\ref{iouteq}) operators filtered about the $\omega_1$ and $\omega_2$ normal mode frequencies \cite{wilson2011}. While achieving electron temperatures in microwave systems well below $70~{\mathrm{mK}}$ is non-trivial, it is feasible; for instance, in Ref. \cite{chang2018}  a calibrated electron temperature of $30~{\mathrm{mK}}$ was achieved.

\section{\label{conclusion}Conclusion}
Motivated by the goal to demonstrate the Unruh effect in tabletop setups, we introduced a model of a pointlike photon detector with its center of mass undergoing oscillatory accelerating motion inside a high quality factor cavity; the detector's internal degrees of freedom are modeled as a quantum harmonic oscillator that is linearly coupled to a massless scalar field in the single mode approximation. Under the condition that the sum of the detector and cavity mode frequencies match that of the detector's center of mass frequency, cavity-detector photon pair production from the vacuum is resonantly enhanced, and the steady state photon production dynamics is accurately described by a simpler non-degenerate parametric amplifier (NDPA) model.
In particular, we derived accurate analytical expressions for the average photon numbers of the detector and cavity modes, as well as the entanglement (logarithmic negativity) between them. The ratio of the coupling strength and the modes' damping rate, denoted as $\eta$, is the determining parameter in the analytical expressions for the average photon numbers and the entanglement; both quantities can be increased by tuning $\eta$ close to $1/2$, the onset of parametric instability.

We proposed an Unruh effect (UE) analogue of the cavity-coupled oscillating detector model, which involves two capacitively coupled co-planar microwave resonators--one playing the role of the cavity and the other the detector. Dilatational vibrations of the coupling capacitance with frequency ($\sim 10\,{\mathrm{GHz}}$) matching the sum of the cavity and detector's fundamental resonator mode frequencies results in measurable,  resonantly enhanced photon production from their ground (i.e., vacuum) state. A key advantage of the analogue scheme is the ability to increase the coupling between the detector and cavity by scaling the capacitance size. 

A complementary tabletop realization of the UE might involve instead scaling up the number of photodetectors, especially if the latter are furnished by atomic scale defects where the individual coupling strengths to the electromagnetic field are fixed by the defect's dimensions. One might consider a large number of such photodetectors embedded for example in a vibrating membrane such that they are all oscillating in unison, the membrane contained within a high quality factor microwave cavity. An interesting question concerns whether a sufficient number of detector defects can be employed such that we enter  a superradiant phase, resulting in a coherent enhancement of the photon production rate from the cavity vacuum \cite{Wang}.

\section{\label{sec:acknowledgements}Acknowledgements}
We thank Oliver B. Wright and Clivia M. Sotomayor-Torres for very helpful discussions. This work was supported by the NSF under Grants No. DMR-1507383, DMR-1507400 and DMR-1807785, and by the ARO under Grant No. W911NF-13-1-0377. CMW acknowledges support from NSERC of Canada and the Canada First Research Excellence Fund.

\appendix
\section{\label{appendixa}The RWA Hamiltonian for the single and many detector case}
For the single-detector case with Hamiltonian given by Eq.~(\ref{orihamiltonian}), applying the Fourier series expansion to $d\tau/dt$ and the Jacobi-Anger expansion to the $\sin\left[{\xi}\cos(\Omega_mt+\phi)/{\Omega_m}\right]$ term, we obtain
\begin{equation}
\frac{d\tau}{dt}=\sum_{n=0}^{\infty}(-1)^n\dbinom{\frac{1}{2}}{n}\dbinom{2n}{n}\left(\frac{\xi}{2}\right)^{2n}+2\sum_{n=1}^{\infty}\sum_{n^{\prime}=1}^n(-1)^{n-n^{\prime}}\dbinom{\frac{1}{2}}{n}\dbinom{2n}{n-n^{\prime}}\left(\frac{\xi}{2}\right)^{2n}\cos\left[2n^{\prime}(\Omega_mt+\phi)\right],
\label{dtaudt}
\end{equation}
\begin{equation}
\sin\left[\frac{\xi}{\Omega_m}\cos(\Omega_mt+\phi)\right]= 2\sum_{n=0}^{\infty}(-1)^nJ_{2n+1}\left(\frac{\xi}{\Omega_m}\right)\cos\left[(2n+1)(\Omega_mt+\phi)\right].
\label{sintermeq}
\end{equation}
Keeping only terms up to second harmonics in $\Omega_m$, Eqs. (\ref{dtaudt}) and (\ref{sintermeq}) become approximately
\begin{eqnarray}
&&\frac{d\tau}{dt}\approx D_0+D_2\cos(2\Omega_mt+2\phi)\\
&&\frac{d\tau}{dt}\sin\left[\frac{\xi}{\Omega_m}\cos(\Omega_mt+\phi)\right]\approx C_1\cos(\Omega_mt+\phi)
\end{eqnarray}
where the $\xi$ dependent $D_0$ and $D_2$ coefficients can be read off from Eq. (\ref{dtaudt}) and the  $\xi$, $\Omega_m$ dependent coefficient $C_1$ can be read off from Eq. (\ref{sintermeq}). Setting the phase $\phi=0$, the Hamiltonian then reduces to
\begin{equation}
H=a^{\dag}a+\left[\omega_d+\omega_{d 0}D_2\cos(2\Omega_mt)\right]b^{\dag}b + \lambda_0 C_1\cos(\Omega_m t)(a^{\dag}+a)(b^{\dag}+b),
\label{seriesapproxhameq}
\end{equation}
where $\omega_d=\omega_{d0} D_0$ is the renormalized detector oscillator frequency. Transforming to the rotating frame via the unitary operator $U_{\mathrm{RF}}(t)=\exp\left(ia^{\dag}at+ib^{\dag}b\left[\omega_dt+\frac{\omega_{d0}D_2}{2\Omega_m}\sin(2\Omega_mt)\right]\right)$, the cavity mode and detector annihilation operators pick up time-dependent phase terms as follows: 
\begin{eqnarray}
a(t)&\to& e^{-it}a(t),\cr 
b(t)&\to& e^{-i\left[\omega_dt+\frac{\omega_{d0}D_2}{2\Omega_m}\sin(2\Omega_m t)\right]}b(t).
\end{eqnarray}
The system Hamiltonian (\ref{seriesapproxhameq}) then becomes  in the interaction picture
\begin{equation}
H_I=\lambda_0C_1\cos(\Omega_mt)(e^{it}a^{\dag}+e^{-it}a)\left[e^{i\omega_dt}e^{iB\sin(2\Omega_mt)}b^{\dag}+e^{-i\omega_dt}e^{-iB\sin(2\Omega_mt)}b\right]
\label{interaction}
\end{equation}
where $B=\omega_{d0}D_2/2\Omega_m<1$.\\
Making use of the Jacobi-Anger expansion again such that
\begin{equation}
e^{\pm iB\sin(2\Omega_mt)}\approx J_0\left(B\right)\pm 2i J_1\left(B\right)\sin(2\Omega_mt),
\end{equation}
where $J_0(z)$ and $J_1(z)$ are Bessel functions of the first kind,  and substituting back to Eq.~(\ref{interaction}),  we arrive at the following expression for the system Hamiltonian:
\begin{eqnarray}
H_I & \approx & \lambda_0C_1\cos(\Omega_mt)\left\{e^{i(\omega_d+1)t}\left[J_0\left(B\right)+2i J_1\left(B\right)\sin(2\Omega_mt)\right]a^{\dag}b^{\dag}\right. \cr
&&\left.+e^{-i(\omega_d+1)t}\left[J_0\left(B\right)-2i J_1\left(B\right)\sin(2\Omega_mt)\right]ab\right. \cr
&&\left.+e^{-i(\omega_d-1)t}\left[J_0\left(B\right)-2i J_1\left(B\right)\sin(2\Omega_mt)\right]a^{\dag}b\right. \cr
&&\left.+e^{i(\omega_d-1)t}\left[J_0\left(B\right)+2i J_1\left(B\right)\sin(2\Omega_mt)\right]ab^{\dag}
\right\}.
\label{interaction2}
\end{eqnarray}
Imposing the parametric resonance condition $\Omega_m=1+\omega_d$ and combining the $\cos(\Omega_mt)$ term with the first two terms within the braces, we  obtain time-independent terms which we retain and oscillating terms at integer multiples of $\Omega_{m}$ which we drop (RWA). 
The resulting, approximate time independent Hamiltonian describes a non-degenerate parametric amplifier (NDPA):
\begin{eqnarray}
H_I & \approx & \frac{\lambda_0C_1}{2}\left[J_0\left(B\right)-J_1\left(B\right)\right](a^{\dag}b^{\dag}+ab).
\end{eqnarray}

For the $N>1$ detectors case with Hamiltonian Eq.~(\ref{manydetectors}), applying the same harmonic expansion approximation as above for the single detector case, we obtain the following approximate Hamiltonian [c.f., Eq. (\ref{seriesapproxhameq})]:
\begin{equation}
H=a^{\dag}a+\sum_{n=1}^N\left[\omega_d+\omega_{d0}D_2\cos(2\Omega_{m}t+2\phi_n)\right]b_n^{\dag}b_n+\lambda C_1\sum_{n=1}^N\cos(\Omega_{m}t+\phi_n)(a^{\dag}+a)(b_n^{\dag}+b_n).
\end{equation}
Transforming to the rotating frame via the unitary operator $\displaystyle U_{\mathrm{RF}}(t)=\exp\left(ia^{\dag}at+i\sum_{n=1}^{N} b_n^{\dag}b_n\left[\omega_dt+\frac{\omega_{d0}D_2}{2\Omega_{m}}\sin(2\Omega_{m}t+2\phi_n)\right]\right)$, the cavity mode and detector annihilation operators pick up time-dependent phase terms as follows: 
\begin{eqnarray}
a(t)&\to& e^{-it}a(t),\cr 
b_n(t)&\to& e^{-i\left[\omega_dt+\frac{\omega_{d0}D_2}{2\Omega_{m}}\sin(2\Omega_{m}t+2\phi_n)\right]}b_n(t).
\end{eqnarray}
Performing again the Jacobi-Anger expansion, imposing the resonance condition $\Omega_{m}=1+\omega_d$ and the RWA, the system Hamiltonian reduces approximately to
\begin{eqnarray}
H_I & \approx & \frac{\lambda_0C_1}{2}\left[J_0\left(B\right)-J_1\left(B\right)\right]\sum_{n=1}^N(a^{\dag}b_n^{\dag}+ab_n).
\label{manydetecthi}
\end{eqnarray}
Note that the phases $\phi_n$ drop out, so that as long as the Hamiltonian (\ref{manydetecthi}) accurately describes the full quantum dynamics (\ref{manydetectors}), there should be little dependence on the relative detector oscillation phases.

\section{\label{appendixb}Analytical derivation of the second order moments}
We start from the Langevin equation~(\ref{ode}) with the input noise operators $a_\mathrm{in}(t)$, $b_\mathrm{in}(t)$ satisfying the expectation value and correlation relations $\langle a_{\mathrm{in}}(t)\rangle=0$, $\langle b_{\mathrm{in}}(t)\rangle=0$ $\langle a_{\mathrm{in}}(t) a_{\mathrm{in}}(t')\rangle=0$, $\langle b_{\mathrm{in}}(t) b_{\mathrm{in}}(t')\rangle=0$, $\langle a^{\dag}_{\mathrm{in}}(t) a_{\mathrm{in}}(t')\rangle=0$, $\langle b^{\dag}_{\mathrm{in}}(t) b_{\mathrm{in}}(t')\rangle=0$  and  $\langle a_{\mathrm{in}}(t)a_{\mathrm{in}}^{\dag}(t^{\prime})\rangle=\delta(t-t^{\prime})$, $\langle b_{\mathrm{in}}(t)b_{\mathrm{in}}^{\dag}(t^{\prime})\rangle=\delta(t-t^{\prime})$. The following linear differential equation results for the second order moments:
\begin{equation}
\frac{d\vec{V}(t)}{dt}=M(t)\vec{V}(t)+\vec{K},
\label{linearode}
\end{equation}
where
\begin{equation}
\vec{V}=
\begin{pmatrix}
\langle aa\rangle \\ \langle a^{\dagger}a\rangle \\ \langle a^{\dagger}a^{\dagger}\rangle \\ \langle ab\rangle \\ \langle a^{\dagger}b\rangle \\ \langle ab^{\dagger}\rangle \\ \langle a^{\dagger}b^{\dagger}\rangle \\ \langle bb\rangle \\ \langle b^{\dagger}b\rangle \\ \langle b^{\dagger}b^{\dagger}\rangle \\
\end{pmatrix},
\vec{K}=\begin{pmatrix}
0 \\ 0 \\ 0 \\ i\lambda \\ 0 \\ 0 \\ -i\lambda \\ 0 \\ 0 \\0 \\
\end{pmatrix}
\end{equation}
and
\begin{equation}
M=
\begin{pmatrix}
-\gamma& 0 &0 &0 &0 &2i\lambda &0 &0 &0 &0\\
0 & -\gamma &0 &-i\lambda &0 &0 &i\lambda &0 &0 &0\\
0 &0 & -\gamma &0 &-2i\lambda &0 &0 &0 &0 &0\\
0 &i\lambda &0 & -\gamma &0 &0 &0 &0 &i\lambda &0\\
0 &0 &-\lambda &0 & -\gamma &0 &0 &-i\lambda &0 &0 \\
-i\lambda &0 &0 &0 &0 & -\gamma &0 &0 &0 &i\lambda\\
0 &-i\lambda &0 &0 &0 &0 & -\gamma &0 &-i\lambda &0\\
0 & 0 &0 &0 &2i\lambda &0 &0 &-\gamma &0 &0\\
0 & 0 &0 &-i\lambda &0 &0 &i\lambda &0 &-\gamma &0\\
0 & 0 &0 & 0 &0 &-2i\lambda &0 &0 &0 &-\gamma
\end{pmatrix}.
\end{equation}

In $\vec{V}$ we include only moments of normal ordered operators, since the moments of anti-normal ordered operators can be obtained from the former via commutation relation identities. From Eq.~(\ref{linearode}) we derive the analytical solution $\vec{V}(t)=\int_0^tdt^{\prime}e^{M(t-t^{\prime})}\vec{K}+e^{Mt}\vec{V}(0)$. With $\vec{V}(0)=0$ (because of normal ordering) and $M$ a nonsingular (i.e., invertible) matrix, we can simplify the form of $\vec{V}(t)$ as follows:
\begin{eqnarray}
\vec{V}(t)&=&\int_0^tdt^{\prime}e^{M(t-t^{\prime})}\vec{K} \cr
&=&\int_0^tdt^{\prime}e^{Mt^{\prime}}\vec{K} \cr
&=&M^{-1}(e^{Mt}-I)\vec{K}.
\end{eqnarray} 
The nonzero elements of $\vec{V}(t)$ are 
\begin{eqnarray}
\langle a^{\dag}(t)a(t)\rangle = \langle b^{\dag}(t)b(t)\rangle = -\frac{\lambda e^{-t (\gamma +2\lambda)} \left[\gamma  \left(e^{4\lambda t}-1\right)+2\lambda\left(-2 e^{t (\gamma +2\lambda )}+e^{4\lambda t}+1\right)\right]}{2(\gamma^2-4\lambda^2)},\label{tdepsecondmom1eq} \\
\langle a^{\dag}(t)b^{\dag}(t)\rangle =-\langle a(t)b(t)\rangle=\frac{i\lambda e^{-t(\gamma +2\lambda)} \left[\gamma\left(-2 e^{t (\gamma +2\lambda)}+e^{4\lambda  t}+1\right)+2\lambda\left(e^{4 \lambda t}-1\right)\right]}{2(\gamma^2-4\lambda^2)},
\label{tdepsecondmom2eq}
\end{eqnarray}
which in the long-time limit reduce to
\begin{eqnarray}
\left.\langle a^{\dag}(t)a(t)\rangle\right|_{t\rightarrow\infty} &=&\left.\langle b^{\dag}(t)b(t)\rangle\right|_{t\rightarrow\infty}=\frac{2\lambda^2}{\gamma^2-4\lambda^2}\cr
\left.\langle a^{\dag}(t)b^{\dag}(t)\rangle\right|_{t\rightarrow\infty} &=& -\left.\langle a(t)b(t)\rangle\right|_{t\rightarrow\infty}=\frac{i\gamma\lambda}{\gamma^2-4\lambda^2}.
\label{secondmoments}
\end{eqnarray}
Equations (\ref{secondmoments}) can also be directly obtained from $\vec{V}(\infty)=-M^{-1}\vec{K}$.

\section{\label{modemodelsec}Derivation of the analogue cavity mode-detector Hamiltonian}
Applying Kirchhoff's laws (currents entering equals currents exiting a node; voltages around a closed loop add to zero), the circuit in Fig.~\ref{schemefig} yields the following equations in terms the cavity and detector flux field variables $\Phi_c(x_c,t)$ and $\Phi_d (x_d,t)$, respectively:
\begin{equation}
\left({\mathcal{C}}+{\mathcal{C}}_{m}\right) \ddot{\Phi}_c -{\mathcal{L}}^{-1} \Phi_c''-{\mathcal{C}}_{m}\ddot{\Phi}_d={\mathcal{C}}_{m}\frac{\partial}{\partial t}\left(\frac{z(t)}{D} \dot{\Phi}_c\right)-{\mathcal{C}}_{m}\frac{\partial}{\partial t}\left(\frac{z(t)}{D} \dot{\Phi}_d\right),\quad c\leftrightarrow d,
\label{circuiteq}
\end{equation}
where the overdots denote partial time derivatives and the prime superscripts denote partial spatial (i.e., $x$-coordinate) derivatives, $z(t)=A\cos(\Omega_m t)$ is the driven dilatational displacement of the FBAR (with $A$ the amplitude and $\Omega_m$ the oscillation frequency), and $D$ is the FBAR thickness. The origin of the $x_c$-coordinate for the cavity center conductor is located at the end that capacitively couples to its probe transmission line. Similarly, the origin of the $x_d$-coordinate for the detector center conductor is located at the opposite end that capacitively couples to its particular probe transmission line. In the overlap region, the cavity and detector coordinates are related as follows: $x_c=x_d+L_c-L_d$.  For notational simplicity, we will frequently drop the $d$ and $c$ subscripts on the $x$ coordinate where the presence of these subscripts already on the flux field variables render these additional subscripts unnecessary.
The parameter ${\mathcal{C}}_{m}$ denotes the undisplaced FBAR capacitance per unit length, which we assume to be well approximated by the parallel plate capacitance formula and we also include a step function in its definition so that
\begin{equation}
{\mathcal{C}}_{m}(x_c)=
\begin{cases}
      0 & 0\leq x_c < L_c-L_{m} \\
      {\mathcal{C}}_{m} &  L_c-L_{m}\leq x\leq L_c,\quad c\leftrightarrow d.
\end{cases}
\label{fbarcapeq}
\end{equation}
From Eq. (\ref{fbarcapeq}), we see that the cavity and detector field variables couple only over the FBAR capacitance length $L_{m}$; for $0\leq x_{c(d)}<L_{c(d)}-L_{m}$, Eq. (\ref{circuiteq}) reduces locally to separate decoupled scalar wave equations in the flux fields  $\Phi_c$ and $\Phi_d$.

Equations (\ref{circuiteq}) follow from the Lagrangian
\begin{eqnarray}
&&L\left[\Phi_d,\Phi_c,\dot{\Phi}_d,\dot{\Phi}_c\right]=\int_0^{L_d} dx\left\{ \frac{1}{2}\left[{\mathcal{C}}+\left(1-\frac{z(t)}{D}\right){\mathcal{C}}_{m}\right] \dot{\Phi}^2_d-\frac{1}{2 {\mathcal{L}}} \Phi_d'^2\right\}  \cr
&&+\int_0^{L_c} dx\left\{\frac{1}{2}\left[{\mathcal{C}}+\left(1-\frac{z(t)}{D}\right){\mathcal{C}}_{m}\right] \dot{\Phi}^2_c-\frac{1}{2 {\mathcal{L}}} \Phi_c'^2\right\}\cr 
&&-\left(1-\frac{z(t)}{D}\right){\mathcal{C}}_{m}\int_0^{L_{m}}dx\,\dot{\Phi}_d \left(L_d-L_{m}+x,t\right) \dot{\Phi}_c\left(L_c-L_{m}+x,t\right).
\label{circuitlagrangeq}
\end{eqnarray}
Here we treat the cavity and detector as closed systems (i.e., neglecting the capacitively coupled transmission lines and other sources of cavity loss), Eq. (\ref{circuiteq}) is accompanied by the following boundary conditions on the flux field variables:
\begin{eqnarray}
\Phi'_c(0,t)&=&\Phi'_c(L_c,t)=0,\quad c\leftrightarrow d.
\label{bceq}
\end{eqnarray}

Noting that the achievable dilatational mode displacement amplitudes satisfy $A\ll D$, we can find approximate solutions to Eq. (\ref{circuiteq}) using the eigenfunction expansion method:  \begin{equation}
\Phi_{c} (x,t)  = \sum_n q_n (t) \Phi_{c, n} (x)\;,\quad c\leftrightarrow d,
\label{efunctionexpeq}
\end{equation}
where the $q_{n}(t)$'s are the normal mode coordinates  labeled by $n=1,2,\dots$, and the associated functions $\Phi_{d, n } (x)$ and $\Phi_{c, n }(x)$ are the normal mode  solutions to the following equations which neglect  the time-dependent oscillatory displacement terms on the right hand side of the equals sign in Eq. (\ref{circuiteq}):
\begin{eqnarray}
\left({\mathcal{C}}_c+{\mathcal{C}}_{m}\right) \omega^2 {\Phi}_c +{\mathcal{C}}_{m}\omega^2{\Phi}_d +{\mathcal{L}}^{-1} \Phi_c''&=&0\;,\quad c\leftrightarrow d\;.
\label{approxcircuiteq}
\end{eqnarray}
From the mode equations (\ref{approxcircuiteq}) and the boundary conditions (\ref{bceq}),  it follows that the normal mode functions satisfy the orthogonality condition
\begin{eqnarray}
&&\int_0^{L_d} dx\left({\mathcal{C}}+{\mathcal{C}}_{m}\right)\Phi_{d,  n'} (x)\Phi_{d,  n} (x) +\int_0^{L_c} dx\left({\mathcal{C}}+{\mathcal{C}}_{m}\right)\Phi_{c,  n'} (x)\Phi_{c,  n} (x)\cr 
&&-{\mathcal{C}}_{m}\int_0^{L_{m}} dx \left[\Phi_{d,  n'} (L_d-x)\Phi_{c,  n} (L_c-x)+\Phi_{c,  n'} (L_c -x)\Phi_{d,  n} (L_d-x)\right]\cr &&=\left(\frac{\Phi_0}{2\pi}\right)^2 C_{n}\delta_{n,n'},
\label{orthogcondeq}
\end{eqnarray}
where the $C_{n}$'s have the dimensions of capacitance, so that the normal mode coordinates $q_{n}(t)$ are dimensionless.
Substituting the mode decomposition (\ref{efunctionexpeq}) into the Lagrangian (\ref{circuitlagrangeq}), applying the boundary conditions (\ref{approxcircuiteq}) and orthogonality condition (\ref{orthogcondeq}) on the mode functions, we obtain the following mode coordinate Lagrangian:
\begin{equation}
L\left[q_{n}, \dot{q}_{n}\right]=\sum_{n,n'} \left(\frac{\Phi_0}{2\pi}\right)^2 \left[ \frac{1}{2} C_{n} \dot{q}_{n }^2 -\frac{1}{2 L_{n}} q_{n}^2 -\frac{z(t)}{2 D} \lambda_{nn'} \dot{q}_{n'}\right],
\label{coordlageq}
\end{equation} 
where $L_{n}^{-1} = \omega_{n}^2 C_{n}$.
The coupling between the normal modes is given by the following formula: 
\begin{equation}
\lambda_{n n'}=\left(\frac{2\pi}{\Phi_0}\right)^2 {\mathcal{C}}_{m}\int_{0}^{L_{m}} dx \left[\Phi_{d,n}(L_d-x)-\Phi_{c,n}(L_c-x)\right]\left[\Phi_{d,n'}(L_d-x)-\Phi_{c,n'}(L_c-x)\right].
\label{couplingaeq}
\end{equation}

Expressing in terms of creation and annihilation operators, the desired closed system, normal mode Hamiltonian that follows from Eq. (\ref{coordlageq}) is approximately
\begin{equation}
H=\sum_{n} \hbar\omega_{n} {a}^{\dag}_{n} {a}_{n} -\frac{z(t)}{D}\sum_{n,n'} \hbar \lambda_{nn'} \sqrt{\omega_{n}\omega_{n'}} \bigl({a}_{n}-{a}^{\dag}_{n}\bigr)\bigl({a}_{n'}-{a}^{\dag}_{n'}\bigr),
\label{creatannihhamaeq}
\end{equation}
where we exploit the fact that $|z(t)| <A\ll D$, and we have redefined the  coupling as follows:
\begin{equation}
\lambda_{nn'}=\left(\frac{\pi}{\Phi_0}\right)^2 \frac{{\mathcal{C}}_{m}}{\sqrt{C_{n} C_{n'}}}\int_{0}^{L_{m}} dx \left[\Phi_{d,n}(L_d-x)-\Phi_{c,n}(L_c-x)\right]\left[\Phi_{d,n'}(L_d-x)-\Phi_{c,n'}(L_c-x)\right].
\label{couplingpeq}
\end{equation}  
Note that the coupling is dimensionless.

\end{document}